\documentclass[a4paper,fleqn,usenatbib]{mnras}

\usepackage[T1]{fontenc}
\usepackage{ae,aecompl}

\usepackage{graphicx}	
\usepackage{amsmath}	
\usepackage{amssymb}	
\graphicspath{{/home/lsesto/Sesto_2017/Figuras_final/}}

\usepackage{rotating}
\usepackage{lscape}
\usepackage[export]{adjustbox}

\title[Young globular clusters]{Young globular clusters in NGC\,1316}

\author[L. Sesto et al.]{
Leandro A. Sesto,$^{1,2}$\thanks{E-mail: sesto@fcaglp.unlp.edu.ar}
Favio R. Faifer,$^{1,2}$
Analía V. Smith Castelli,$^{2,3}$
\newauthor
Juan C. Forte,$^{3,4}$
Carlos G. Escudero$^{1,2}$
\\
$^{1}$Facultad de Cs. Astron\'omicas y Geof\'isicas, Univ. Nac. de La Plata, Argentina\\
$^{2}$Instituto de Astrof\'isica de La Plata, La Plata, Argentina\\
$^{3}$Consejo Nacional de Investigaciones Cient\'ificas y T\'ecnicas (CONICET), Argentina\\
$^{4}$Instituto Argentino de Matem\'atica Alberto Calder\'on, CONICET, Argentina.
}

\date{Accepted XXX. Received YYY; in original form ZZZ}

\pubyear{2017}

\begin{document}
\label{firstpage}
\pagerange{\pageref{firstpage}--\pageref{lastpage}}
\maketitle

\begin{abstract}

We present multi-object spectroscopy of the inner zone of the globular cluster (GC) system associated with the intermediate-age merger remnant NGC\,1316. Using the multi-object mode of the GMOS camera, we obtained spectra for 35 GCs. We find pieces of evidence that the innermost GCs of NGC\,1316 rotate almost perpendicular to the stellar component of the galaxy.
In a second stage, we determined ages, metallicities and $\alpha$-element abundances for each GC present in the sample, through the measurement of different Lick/IDS indices and their comparison with simple stellar population models.
 We confirmed the existence of multiple GC populations associated with NGC\,1316, where the presence of a dominant subpopulation of very young GCs, with an average age of 2.1 Gyr, metallicities between -0.5~\textless~[Z/H]~\textless~0.5\,dex and $\alpha$-element abundances in the range -0.2~\textless~[$\alpha$/Fe]~\textless~0.3\,dex, stands out. Several objects in our sample present subsolar values of [$\alpha$/Fe] and a large spread of [Z/H] and ages. Some of these objects could actually be stripped nuclei, possibly accreted during minor merger events.
Finally, the results have been analyzed with the aim of describing the different episodes of star formation and thus provide a more complete picture about the evolutionary history of the galaxy. We conclude that these pieces of evidence could indicate that this galaxy has cannibalized one or more gas-rich galaxies, where the last fusion event occurred about 2 Gyr ago.

\end{abstract}




\section{Introduction}
\label{Intro}

Great efforts have been made in the last decades to describe those processes that give rise to the formation of different subpopulations of GCs in different environments. The pioneering works of \citet{Ashman1992, Forbes1997} and \citet{Cote1998} formed the basis for developing more complex scenarios that included a cosmological context \citep[e.g.][]{Beasley2002, Pipino2007, Muratov2010,Kruijssen2015}.
Although there is still no definitive scenario that explains both the formation of globular clusters (GCs) and their most significant properties, it is believed that these objects have been originated in the most intense star formation episodes in the life of galaxies \citep[e.g.][]{Forte2014, Kruijssen2015}. This last work proposes that the origin of ``classical'' GCs could be explained with a two-phase model, where star cluster populations would initially be formed with high efficiency at z~\textgreater~2 due, among other reasons, to the high rates of galaxy mergers at high redshift. This scenario implies that, although the galaxy mergers at z~$\leq$~1 are much rarer,  in the present-day Universe we can observe young star and globular clusters in gas-rich merging galaxies. For this reason GCs have a significant role in several aspects related to the study of stellar evolution and the chemical enrichment that occurred during the main events in the history of galaxy assembly, making them good tracers of the different stages of galaxy formation and evolution.

If GCs are real tracers of the formation of the dominant stellar populations of the galaxies, they are expected to share some common features with field stars, like their chemical abundance, spatial distribution and age. Some pioneering works exploring this idea are those by \citet{Eggen1962}, \citet{Searle1978}, and more recently, it has been explored by, e.g., \citet{Forte2014}. Furthermore, although recent works show the presence of multiple stellar populations in many GCs of our Galaxy \citep[e.g.][]{Bedin2004, Piotto2007, Milone2017}, it should be noted that they are still a very good representation of simple stellar population (SSP) models. These aspects significantly simplify, for example, the determination of their ages and metallicities.

While GC systems associated with early-type galaxies are mainly constituted by old objects with ages in the range of 9$-$13 Gyr, the existence of some young and intermediate-age GCs originate in gas-rich merger events can be found. Few works in the literature address the analysis of this kind of objects \citep[e.g.][]{Goudfrooij2001a, Schweizer2002, Strader2004, Woodley2010a, Park2012, Trancho2014}, however, a complete study of their integrated properties is still missing.

In this context, this work is focused in a detailed analysis of young objects belonging to the GC system associated with NGC\,1316, the brightest galaxy of the Fornax galaxy cluster. This giant elliptical (E) galaxy is one of the closest and most intense radio sources in the southern hemisphere (Fornax A), located on the outskirts of the Fornax cluster to a distance of 20.8 Mpc \citep{Cantiello2013}.

NGC\,1316 displays a number of morphological features which indicate that this galaxy is a merger remnant of approximately 3 Gyr \citep{Goudfrooij2001a}. Among them, we can emphasize shells, ripples \citep{Schweizer1980, Schweizer1981} and an unusual pattern of dust, formed by large filaments and dark structures \citep{Carlqvist2010}. This galaxy has been studied in almost all wavelengths, from X-rays to radio frequencies (see \citealt{Richtler2012b} and \citealt{Iodice2017} for an interesting summary). Table \ref{table0} lists some relevant properties of NGC\,1316.

On the other hand, the GC system of NGC\,1316 has been previously studied by \citet{Gomez2001, Goudfrooij2001a, Goudfrooij2001b, Goudfrooij2004, Richtler2012a, Richtler2012b, Richtler2014, Sesto2016}. Particularly, the paper of \citet{Goudfrooij2001a} deserves special attention, because this is the only work that presents spectroscopic ages and metallicities, but unfortunately only for a sample of three GCs.

In \citet{Sesto2016} (hereafter Paper I), we established photometrically the presence of different GCs subpopulations likely associated with the merger events. In that work, the integrated $(g-i)'$ colour distribution analysis reveals the ``classical'' blue and red GCs with colour peaks at $(g-i)'_0 \sim$ 0.83 and 1.13\,mag respectively, a significant GC subpopulation with intermediate colours ($(g-i)'_0 \sim$ 0.96\,mag), and the possible existence of a fourth group of clusters at $(g-i)'_0 \sim$ 0.42\,mag. In addition, an attempt to derive ages and chemical abundances comparing the photometric results with different SSP models was made. The analysis determined that the modal colours of the intermediate candidates were reasonably well represented by a 5 Gyr isochrone and a solar (or slightly subsolar) metallicity. Throughout this paper, the $(g-i)'_0$ colour ranges adopted to separate between the different GC subpopulations were 0.4$-$0.90 for the blue GCs, 0.95$-$1.05 for the intermediate and 1.05$-$1.4 for the red ones.

In order to constrain the star formation episodes in the galaxy and understand the formation of GCs in NGC\,1316, we carried out a spectroscopic study of 40 GC candidates associated with the galaxy, located within a projected distance of 25 Kpc from the galactic centre.

\begin{figure}
	\includegraphics[width=0.95\columnwidth]{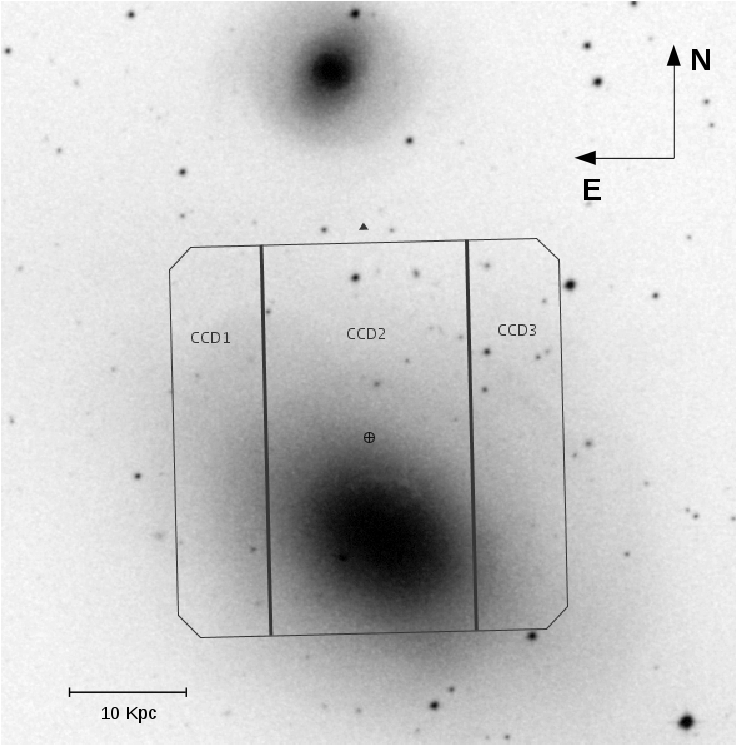}
    \caption{GMOS field used as pre-image for spectroscopy, superposed to the DSS image of NGC\,1316.}
    \label{fig1}
\end{figure}
%
\begin{table}
\centering
\footnotesize
\begin{tabular}{lll}
\hline
\hline
Parameter&Value&Reference\\
\hline
$\alpha_{J2000}$&$3^{h}22^{m}41.7^{s}$&NED\\
$\delta_{J2000}$&$-37\degr12'30''$&NED\\
Type&(R')SAB(s)0 pec&RC3\\
$V^{T}_{0}$&8.53\,mag&RC3\\
$(m-M)_{0}$&31.59$\pm$0.05\,mag&\citet{Cantiello2013}\\
Distance&20.8 Mpc&\citet{Cantiello2013}\\
PA$^\dag$&55\degr$\pm$0.2&\citet{Sesto2016}\\
$V_{helio.}$&1760$\pm$10\,km/s&\citet{Longhetti1998}\\
$\sigma$&240$\pm$11\,km/s&\citet{Longhetti1998}\\
$R_{eff}$&2'.22&\citet{Iodice2017}\\
$A_{V}$&0.057\,mag&\citet{Schlafly2011}\\
$M_{V^{T}_{0}}$&-23.12\,mag&$\bigstar$\\

\hline
\end{tabular}
\caption{Main properties of NGC\,1316. ($\bigstar$) Calculated using the above data. ($\dag$) Corresponds to
the photometric major axis of the galaxy.}
\label{table0}
\end{table}

\section{Observations and Data Reduction}
\label{sec2.1}

The dataset presented in this work was obtained using the multi-object mode of the Gemini Multi-object Spectrograph (GMOS), mounted on the Gemini South telescope. The selection of spectroscopic sources and their photometric data were taken from Paper I. In turn, for the design of the MOS mask, the central field of the photometric mosaic was used as a pre-image (see Figure \ref{fig1}).

As a first step, we selected the slits prioritizing the GCs spectroscopically confirmed by \citet{Goudfrooij2001a} and photometric GC candidates brighter than $V=$ 21.7\,mag ($g'_0 \sim$ 22\,mag). This preferential value was established with the goal of reaching signal-to-noise (S/N) values high enough to obtain ages and metallicities with low errors. In addition, slits were placed on weaker objects, reaching V$_0$ $\sim$ 23\,mag. 

The data were obtained between August 2013 and January 2014, under photometric conditions, as part of Gemini programme GS$-$2013B$-$Q$-$24 (PI: Leandro Sesto). The MOS mask consisted of 40 slits of 1 arcsec width and 4$-$6 arcsec length. We used the B600$-$G5303 grating centred at 5000 and 5100\AA ~(to cover the CCD chip gaps), with 2 $\times$ 2 binning, and exposure times of 16$\times$1800 seconds, yielding 8 hours of on source integration. Table \ref{table1} shows the basic information related to all our data.

In order to perform the individual calibrations, flat-fields and copper-argon (CuAr) arc spectra were observed with each science image. In turn, bias images were obtained from the ``Gemini observatory Archive'' (GOA).\\

The images were processed using the GEMINI-GMOS routines in IRAF (version V2.16). This process was carried out in different stages, which included corrections by bias and flat field, calibration in wavelength, and the extraction and subsequent combination of individual spectra. A brief summary of the steps followed is shown below:\\

\begin{table}
\centering
\footnotesize
\begin{tabular}{lccccc}
\hline
\hline
Object&Date&Grism&Exp.&$\lambda$$_{c}$&FWHM\\
&&&(sec)&(nm)&(arcsec)\\
\hline
NGC\,1316&08 Sep 2013&B600&2$\times$1800&500&0.78\\
&&&2$\times$1800&510&0.83\\
&30 Nov 2013&&2$\times$1800&500&0.62\\
&&&2$\times$1800&510&0.65\\
&26 Dec 2013&&1$\times$1800&510&0.65\\
&28 Dec 2013&&1$\times$1800&510&0.63\\
&31 Dec 2013&&1$\times$1800&500&0.65\\
&&&1$\times$1800&510&0.61\\
&01 Jan 2014&&1$\times$1800&500&0.55\\
&05 Jan 2014&&1$\times$1800&500&0.75\\
&&&1$\times$1800&510&0.70\\
&06 Jan 2014&&1$\times$1800&500&0.73\\
LTT7379&02 Sep 2013&B600&1$\times$10&430&1.45\\
&&&1$\times$10&510&1.30\\
&&&1$\times$10&590&1.40\\
\hline
\end{tabular}
\caption{List of individual exposures belonging to the program GS-2013B-Q-24. The observation program included a flat-field exposition and a CuAr lamp spectrum for each science image.}
\label{table1}
\end{table}
\begin{figure}
	\includegraphics[width=0.9\columnwidth,angle=270]{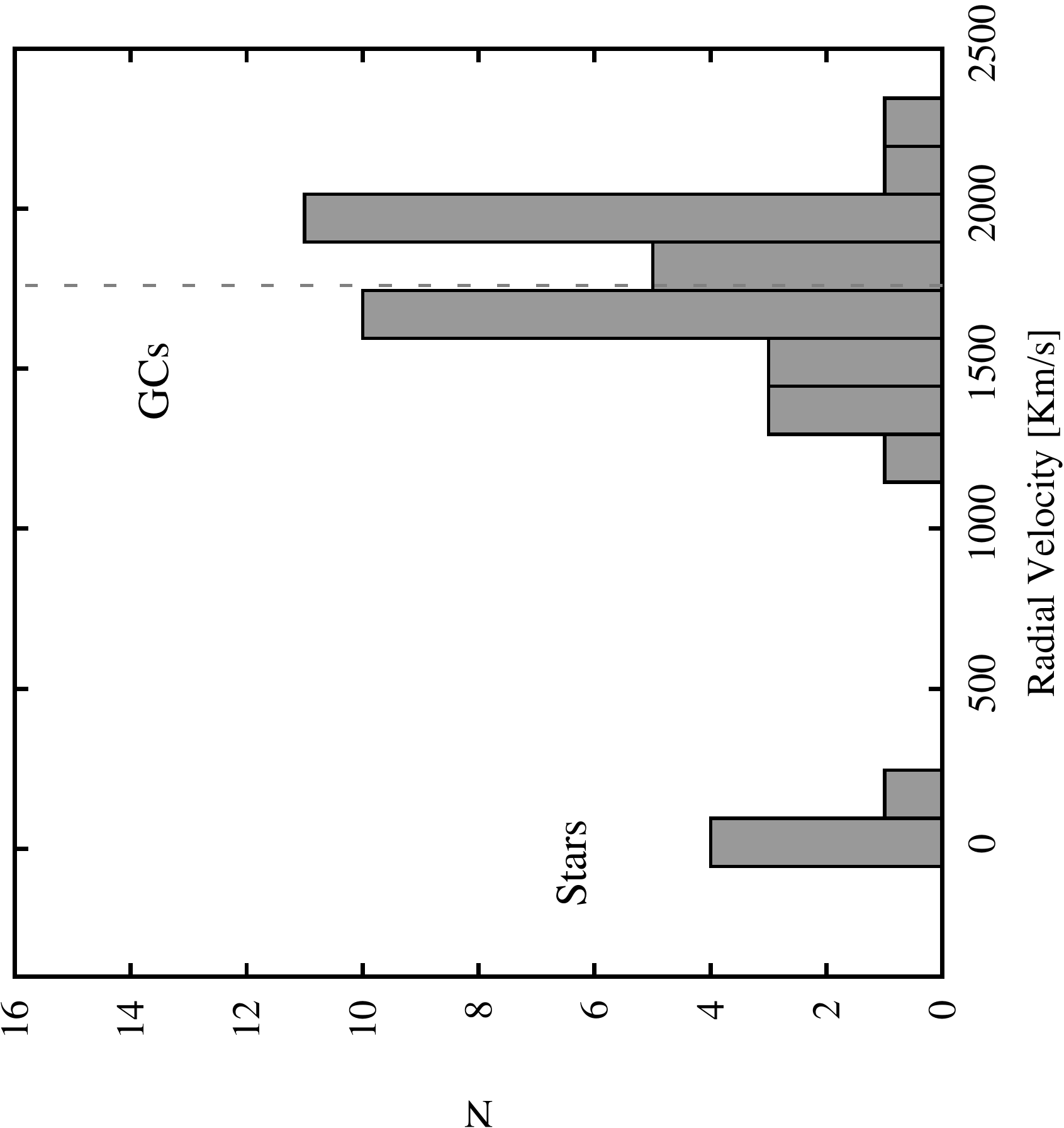}
    \caption{RV distribution diagram. The vertical dashed line indicates the systemic velocity of NGC\,1316.}
    \label{fig2}
\end{figure}

-The task {\sc{gbias}} was used to generate different master-bias for the distinct dates of observation. These were used both for the fit and subtraction of the overscan level, as well as for correction by bias in the images. This last action was developed by the {\sc{gsreduce}} task.

-Science frames  were flat-fielded using the {\sc{gsreduce}} task and the mask of bad pixels was applied through the {\sc{fixpix}} task.

-The cosmic rays present in the images were cleaned using the Laplacian Cosmic Ray Identification\footnote{Laplacian Cosmic Ray Identification routine by P. van Dokkum
(http://www.astro.yale.edu/dokkum/lacosmic/)} routine \citep{Vandokkum2001}.

-Before the wavelength calibration, the individual spectra must be identified to subsequently proceed to cut and to store them in a new multi-extension image. It is essential that both, the science images and their respective calibration images are cut identically. For this reason, a reference image which contains information of the positions of the different slits present in the mask was generated for each data package. In order to carry out this assignment the task {\sc{gscut}} was used to identify the slits in flat-field images without normalization, using an edge-detecting algorithm on the gradients produced at the edges of the slides.

-The task {\sc{gswavelength}} was used to obtain the transformation both to rectify the spectra and to obtain the wavelength calibration from the CuAr arc frames. Calibration was established independently for each spectrum. In all cases, fourth or fifth order fits were made, in which the \textit{rms} obtained did not exceed 0.2 pixels.

-The rectification and wavelength calibration of the science spectra were carried out through the {\sc{gstransform}} task, using the corresponding solutions obtained in the CuAr lamps.

-The task {\sc{apall}} was used to extract the individual spectra and perform subtraction of the sky.

-The 16 extracted spectra of each slit were  normalized according to the signal between 4500 and 5500 \AA. Then, they were combined  to obtain a final median spectrum using the task {\sc{scombine}}. We selected SIGCLIP algorithm for deviating pixel rejection.

In order to obtain a representative value of S/N per\,\AA\,for each spectrum, a simple average of this values between 5000 and 5050\AA\, was made. These values are expressed in the last column of Table \ref{table3}.

Finally, spectroscopic standard star observations were used to transform our instrumental spectra into  approximate flux-calibrated spectra. That is, our data includes the acquisition of long-slit spectra of the standard star LTT7379. These calibration spectra were obtained as part of the observation program, using the same configuration as the science data.

The obtained MOS spectra typically cover the range 3500$-$6500\AA, with a dispersion of 0.90 \AA/pix and a spectral resolution of approximately 4.7\,\AA. Most of them have an excellent S/N (some of them with S/N per\,\AA~\textgreater~50). This allowed us to determine radial velocities, ages, metallicities and $\alpha$-elements abundance for each GC present in the sample.\\

\section{Kinematics}
\label{sec3}

\subsection{Confirmation of GCs}
\label{sec3.1}

Radial velocities (RV) were obtained using the method of cross-correlation \citep{Tonry1979} with IRAF. As template spectra we used SSP models obtained from the MILES libraries \citep{Vazdekis2010}. A total of 19 SSP models were considered, covering a wide range of ages (2.5, 5 and 12.6 Gyr) and metallicities ([Z/H] = -2.32; -1.71; -1.31; -0.71; -0.4 ; 0; 0.4 dex), with a unimodal initial mass function (IMF) with a slope value of 1.3. For each GC candidate present in the mask we obtained 19 estimates of its RV, that is, one for each model. Then these values were averaged by the mean. It should be noted that only those values that were within the 3$\sigma$ of the mean value were averaged.
Our instrumental configuration did not allow us to determine the internal velocity dispersion of the GCs (typical values of the order of 10 km/s, \citealt{Harris2010}).
We did not apply heliocentric corrections because they are smaller than the formal errors of the RV.

\begin{figure}
	\includegraphics[width=0.95\columnwidth,angle=270]{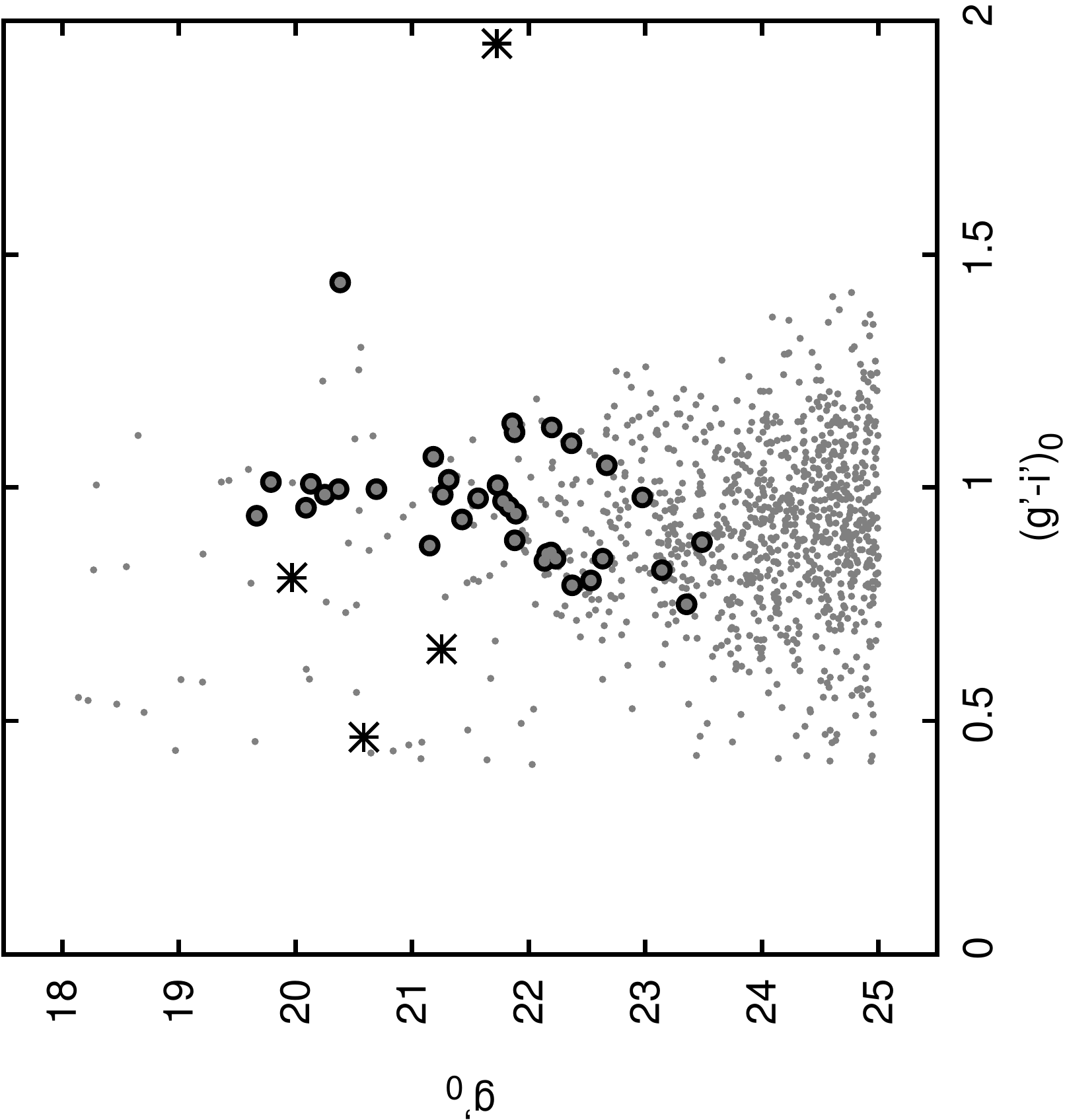}
    \caption{$g'_0$ versus $(g-i)'_0$ colour-magnitude diagram. Point sources with $g'_0$~\textless~25\,mag are shown with small grey dots, spectroscopically confirmed GCs with big circles and field stars with asterisks. Object ID$_S$=351 is observed at the red end of the figure (see Section \ref{sec4.2.0}).}
    \label{fig3n}
\end{figure}
\begin{figure}
	\includegraphics[width=0.95\columnwidth,angle=270]{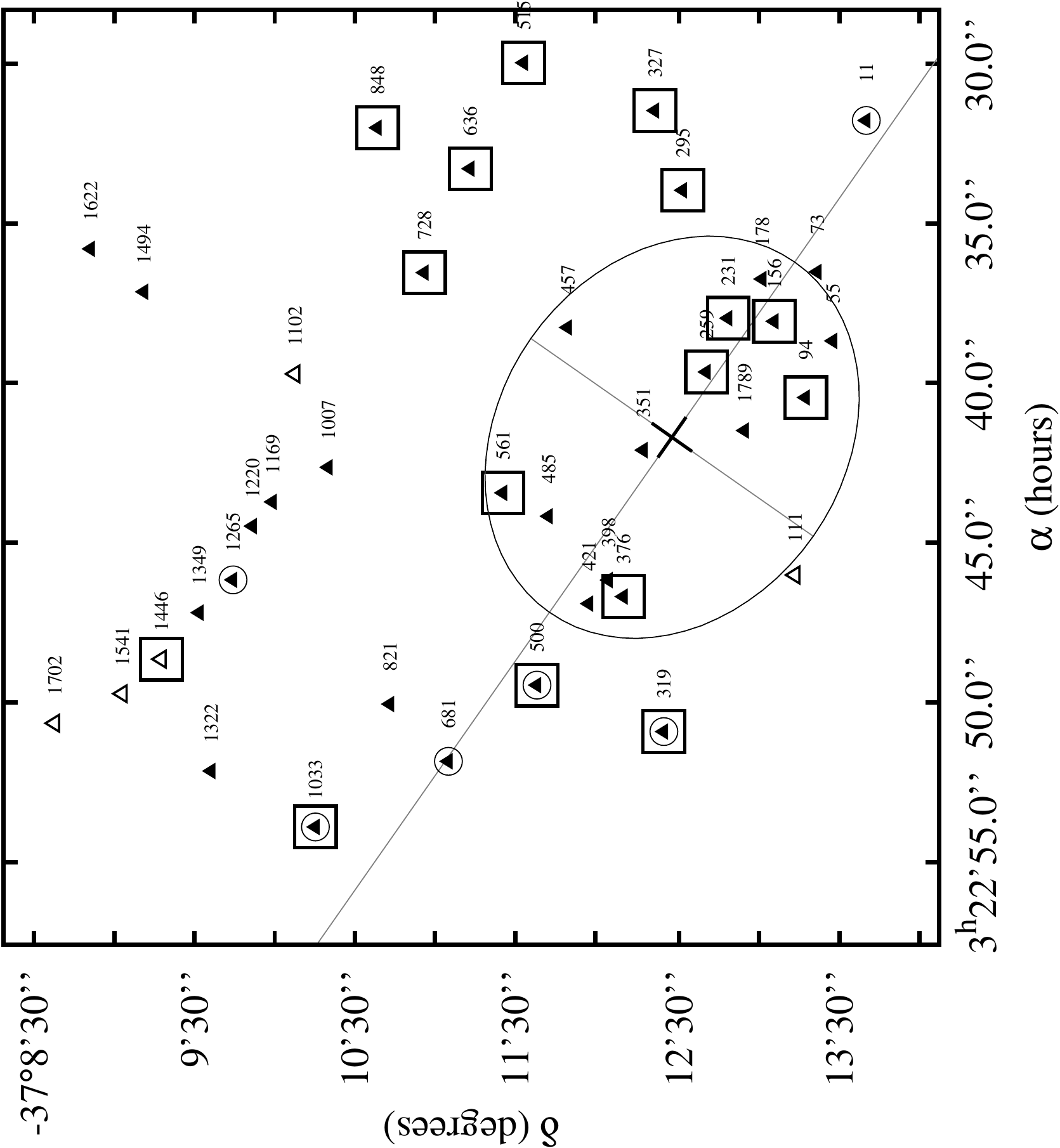}
    \caption{Positions of the GCs candidates belonging to the spectroscopic sample. The centre of NGC\,1316 is displayed with a cross. Confirmed GCs are shown with filled triangles and field stars with open triangles (see Section \ref{sec3.1}). Boxes represents GCs in common with \citet{Goudfrooij2001a} and circles, those in common with \citet{Richtler2014}. The ellipse indicates the position and orientation of the galaxy at $\mu_g$=21.5 mag arcsec$^{-2}$. The solid lines indicate the photometric major and minor axis of the galaxy \citep{Sesto2016}.}
    \label{fig4n}
\end{figure}
\begin{figure}
\centering
	\includegraphics[width=0.9\columnwidth, angle=270]{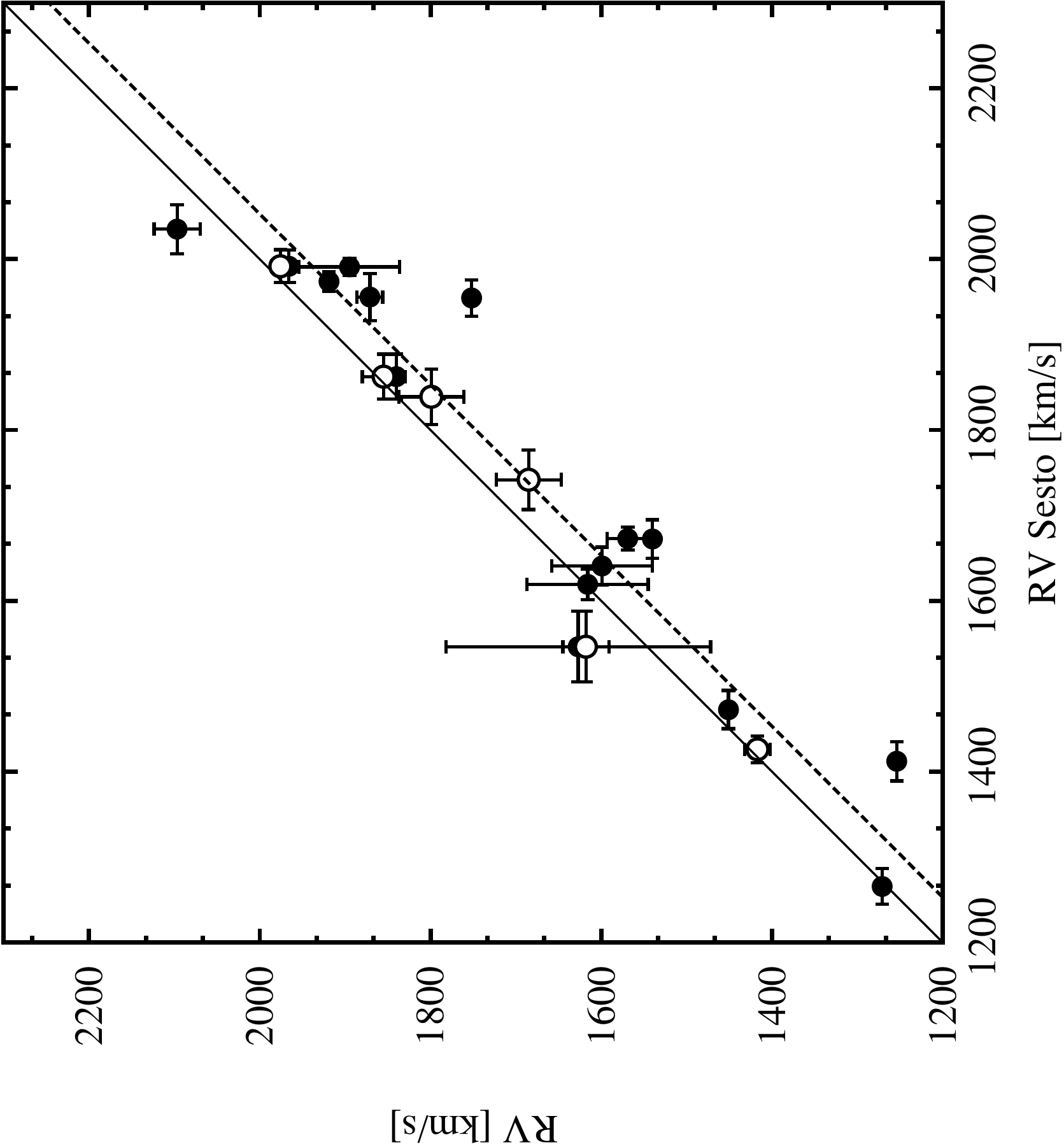}
    \caption{Comparison between RV measured in this work and those obtained by \citet{Goudfrooij2001a} (filled circles) and \citet{Richtler2014} (empty circles). The solid line indicates the line with unitary slope. The dashed line indicates the least-squares fit with \citet{Goudfrooij2001a} data.}
    \label{fig5n}
\end{figure}

Figure \ref{fig2} shows the RV distribution for all objects present in the mask. The size of the adopted bin is 150\,km/s, much higher than the average errors. We adopted 1760\,km/s, as the systemic velocity of NGC\,1316 \citep{Longhetti1998}. The big difference of mean RV between the genuine GCs and the field stars belonging to the Milky Way (MW) is easily appreciated. To quantify this description a Gaussian velocity distribution was adopted for the GC system and those with $|$RV$_{mean}-$RV$|$ $\leq$ 3$\sigma$ were considered as members of NGC\,1316. We obtained a mean RV of 1793$\pm$50\,km/s with a velocity dispersion of 241$\pm$48\,km/s. Thirty-five genuine globular clusters were confirmed and
only 5 objects present in the sample were field stars with heliocentric radial velocities lower than 100 km/s (ID$_S$ $\#$111, 1102, 1446, 1541 and 1702 in Table \ref{table3}).

Figure \ref{fig3n} shows the colour-magnitude diagram (CMD) of the full spectroscopic sample and the unresolved objects of the photometric mosaic presented in Paper I.
One of the field stars, ID$_S$=1102, is outside the range of the figure, since it has colour $(g-i)'_0$ = 2.91\,mag. This object, much redder than the proposed limit for ``classical'' GCs (see Section 3.2 in \citealt{Sesto2016}), was included to occupy an empty space in the mask.
 In the same way the object ID$_S$=351 (which has a colour $(g-i)'_0$=1.44\,mag) appears as an interesting object, because according to its RV constitutes a bona-fide GC. Its location in the diagram can be tentatively explained by considering that it is strongly reddened due to the innermost dusty structure (see Section \ref{sec4.2.0}).

The group of GCs brighter than $g'_0$= 21\,mag is outstanding, since these objects are much brighter than the GCs of the MW. For example, $\omega$Centauri, the brightest GC of the MW, has M$_V$ = -10.26\,mag and $(V-I)_0$ = 1.05\,mag \citep{Harris1996}. At the distance of NGC\,1316 and using transformations [2] and [3] from \citet{Faifer2011}, this value is equivalent to $g'_0$ = 21.6\,mag. Therefore, the brightest GCs associated with NGC\,1316 are at least one magnitude brighter than $\omega$Centauri. It is expected that NGC\,1316  will host several GCs as bright as $\omega$Centauri simply due to a size-of-sample effect. Similar behaviour is observed, by e.g., in the sample of early-type Galaxies studied by \citet{Faifer2011}. On the other hand, if these bright GCs were formed during the last merger events, their high magnitudes ($g'_0$~\textless~21.6\,mag) could be explained simply because they are younger than a typical GC, as can be seen in \citet{Whitmore1997}. That is, GCs of $\sim$3 Gyr tend to be 1$-$2\,mag brighter than GCs of 13 Gyr displaying similar metallicities.


\begin{figure}
	\includegraphics[width=0.9\columnwidth, angle=270]{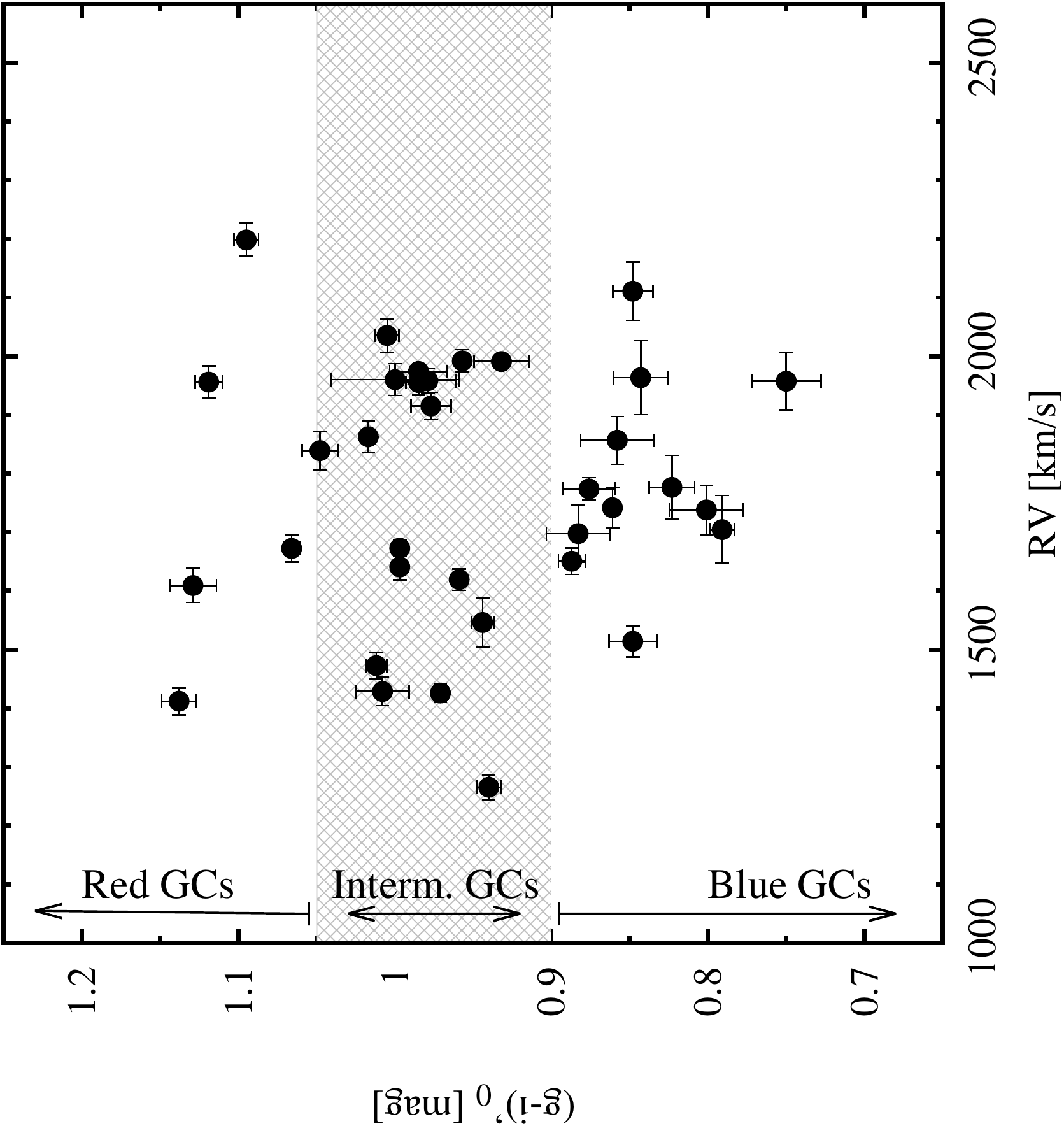}
 \caption{$(g-i)'_0$ colour as a function of the RV for GCs belonging to our sample. The dashed straight line shows the systemic velocity of NGC\,1316 (1760\,km/s). The scratched area indicate the colour ranges for the intermediate-colours GCs.}
    \label{fig7n}
\end{figure}

\subsection{Comparison with previous measurements}
\label{sec3.2}

In the literature several kinematic papers related to NGC\,1316 can be found. Among them, we will pay particular attention to the works of \citet{Goudfrooij2001a} and \citet{Richtler2014}, based on the study of the GCs and the work of \citet{McNeil2012}, which studies the kinematics of the planetary nebulae (PNe) in NGC\,1316.

\begin{figure}
\centering
	\includegraphics[width=0.9\columnwidth, angle=270]{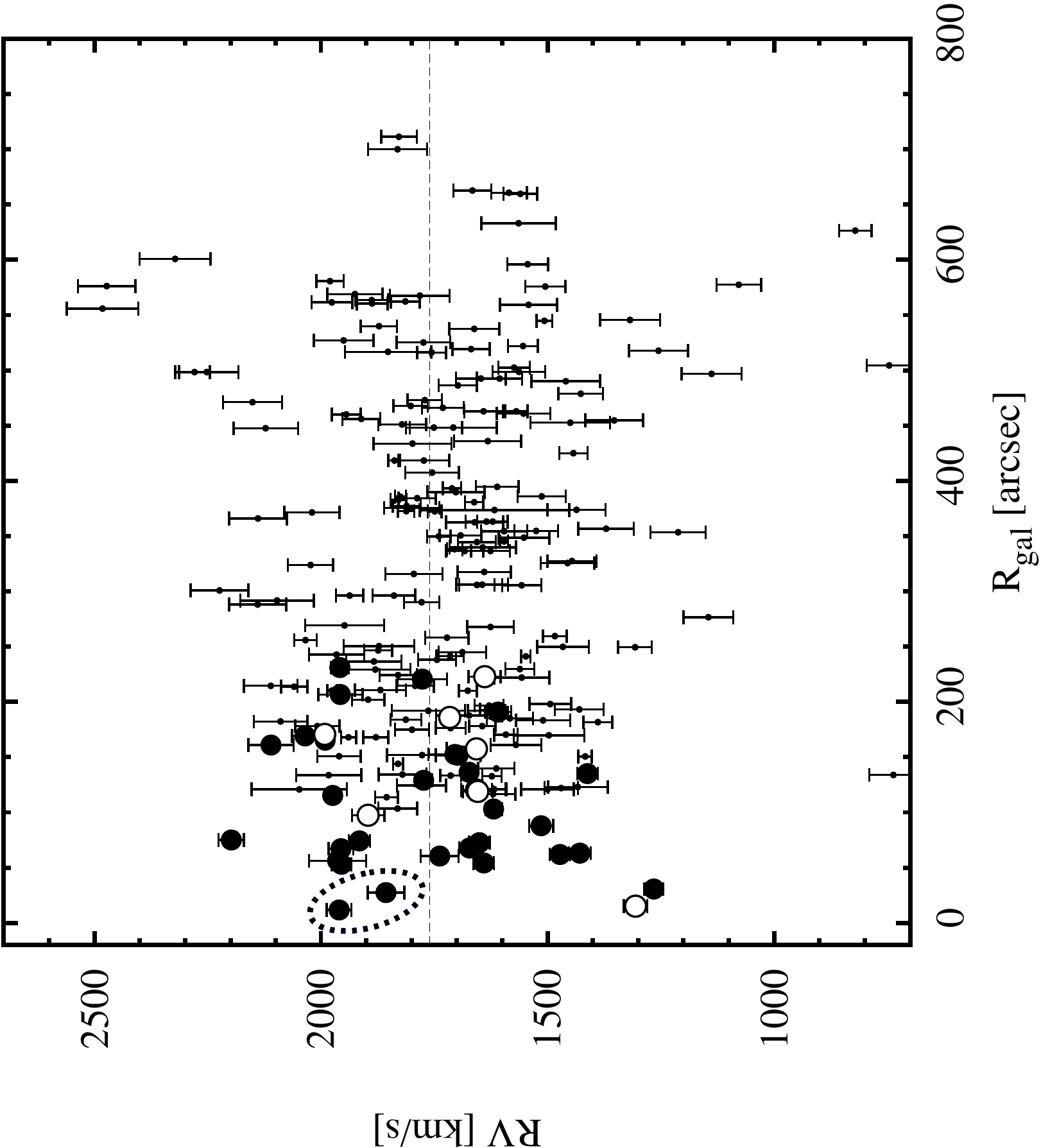}
    \caption{Radial velocity vs galactocentric radii. Large filled circles show our measurements. Small dots represent the objects measured by \citet{Richtler2014}, which extend to distances of $\sim$800 arcsec from the galactic centre. The open circles represent the objects measured by \citet{Goudfrooij2001a}. The dotted straight line shows the systemic velocity of NGC\,1316 (1760 km/s). The dashed ellipse is defined in the text.}
    \label{fig8n}
\end{figure}
\begin{figure*}
\centering
	\includegraphics[width=0.9\columnwidth, angle=270]{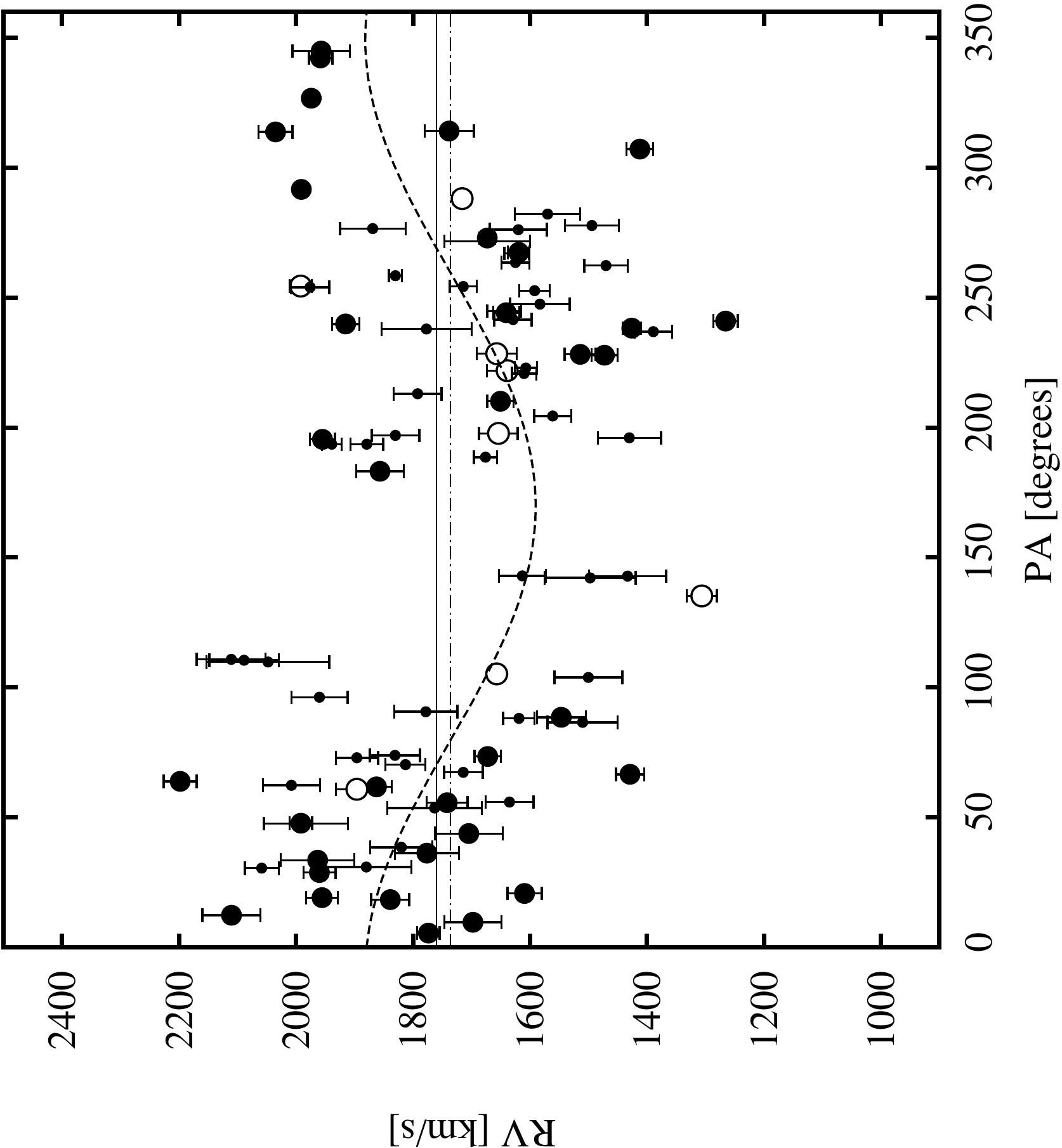}
	\includegraphics[width=0.9\columnwidth, angle=270]{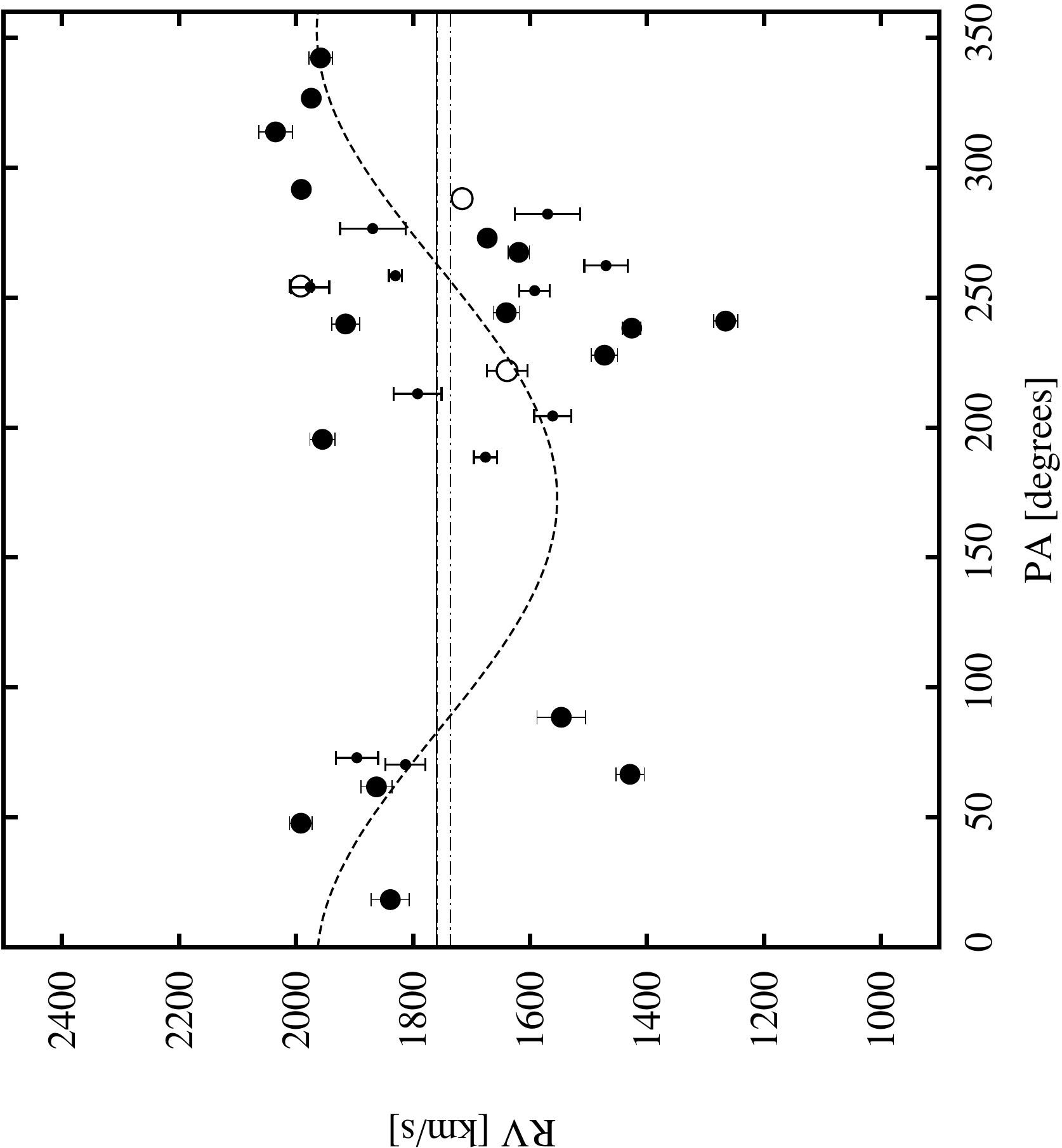}
    \caption{Radial velocity as a function of PA. The GCs samples of \citet{Goudfrooij2001a} (open circles), \citet{Richtler2014} (small filled circles) and of this work (large filled circles) are shown. The dashed line represents the best fit. The continuous straight line shows the systemic velocity of NGC\,1316 (1760 km/s) and the straight dash-dotted line the GCs systemic velocity.}
    \label{fig9n}
\end{figure*}

\citet{Goudfrooij2001a} presented multi-object spectroscopy of 24 GCs, using observations from the New Technology Telescope (NTT). Sixteen GCs of their sample are present in our spectroscopy (see Table \ref{table3}). The only discrepancy between the two datasets is the nature of the object ID$_{Goud.}$=119 (ID$_{Sesto}$=1446). This object, located in the blue limit of our spectroscopic sample ($(g-i)'_0$=0.465\,mag), was catalogued as a genuine GC by \citet{Goudfrooij2001a} who measured a RV of 1970$\pm$60\,km/s. Instead, we estimate a RV~=~-38$\pm$30\,km/s, which would be indicating that it is actually a field star.
These authors also obtained three spectra with enough S/N to establish ages and metallicities. A more detailed discussion of this study will be made in Section \ref{sec4.1}.
In turn, \citet{Richtler2014} present wide-field multi-object spectroscopy for 562 GC candidates, obtained through the FORS2 instrument mounted in the Very Large Telescope (VLT). They confirmed 177 GCs by estimating their RV, of which only six are present in our spectroscopy (see Table \ref{table3}). Unfortunately the spectra presented by \citet{Richtler2014} did not have enough S/N to obtain ages and metalicities. Figure \ref{fig4n} shows the spatial distribution of objects in common with the different spectroscopic samples.

Figure \ref{fig5n} shows the comparison between RVs measured in this work and those obtained by \citet{Goudfrooij2001a} and \citet{Richtler2014} for those GCs in common.
In the figure it is observed that there seems to be a systematic offset between these samples and ours. A tentative correction of RV values was carried out in order to obtain a homogeneous sample  with our data. Through a least-squares fit the best-fit zero points are determined. For the sample of \citet{Goudfrooij2001a} we apply a zero point correction of 53$\pm$20 km/s. However, for the sample of \citet{Richtler2014} this value was lower than 10 km/s, so we decided not to apply corrections to the radial velocities of these authors.

\begin{figure*}
\centering
	\includegraphics[width=0.9\columnwidth, angle=270]{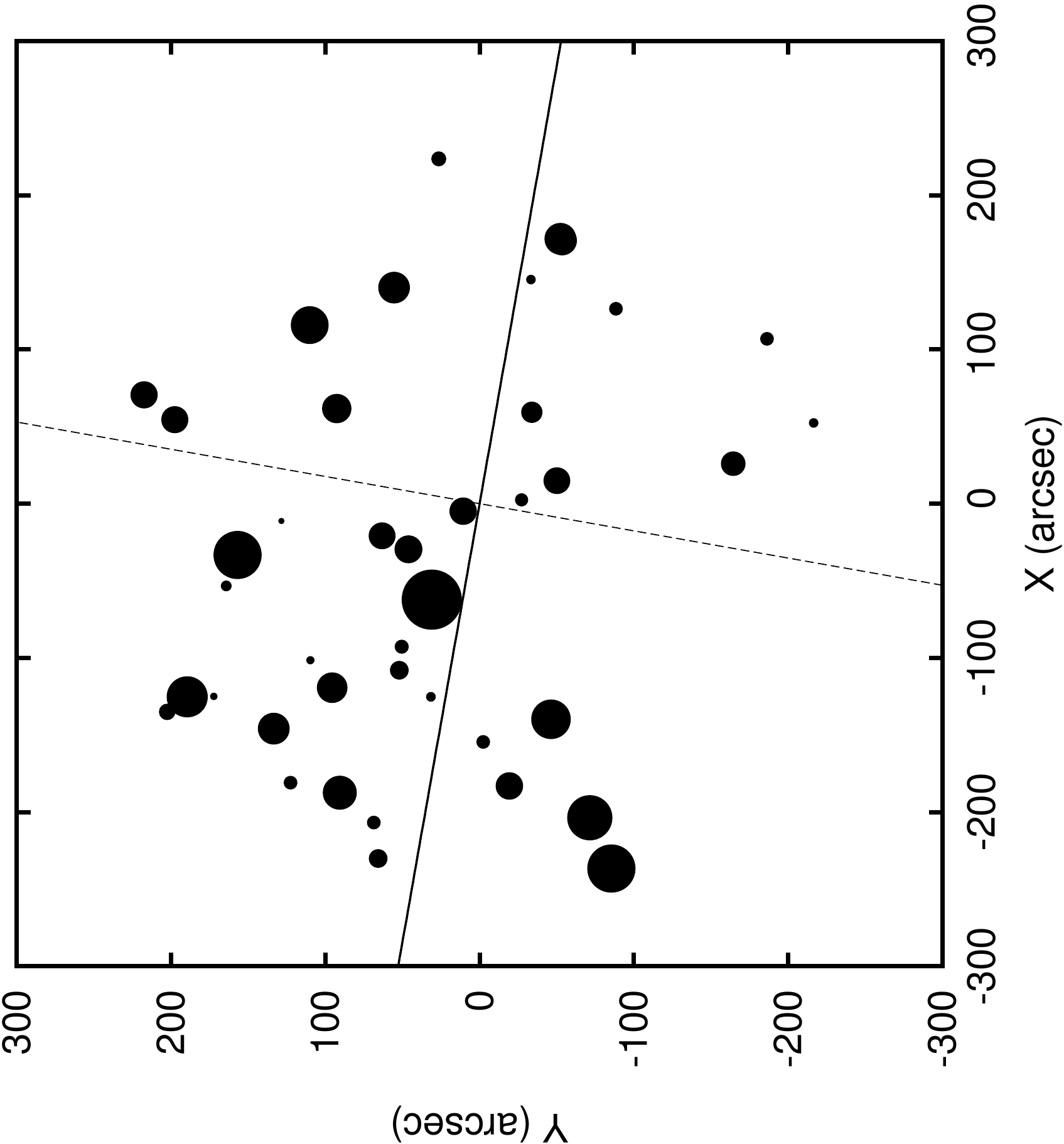}
	\includegraphics[width=0.9\columnwidth, angle=270]{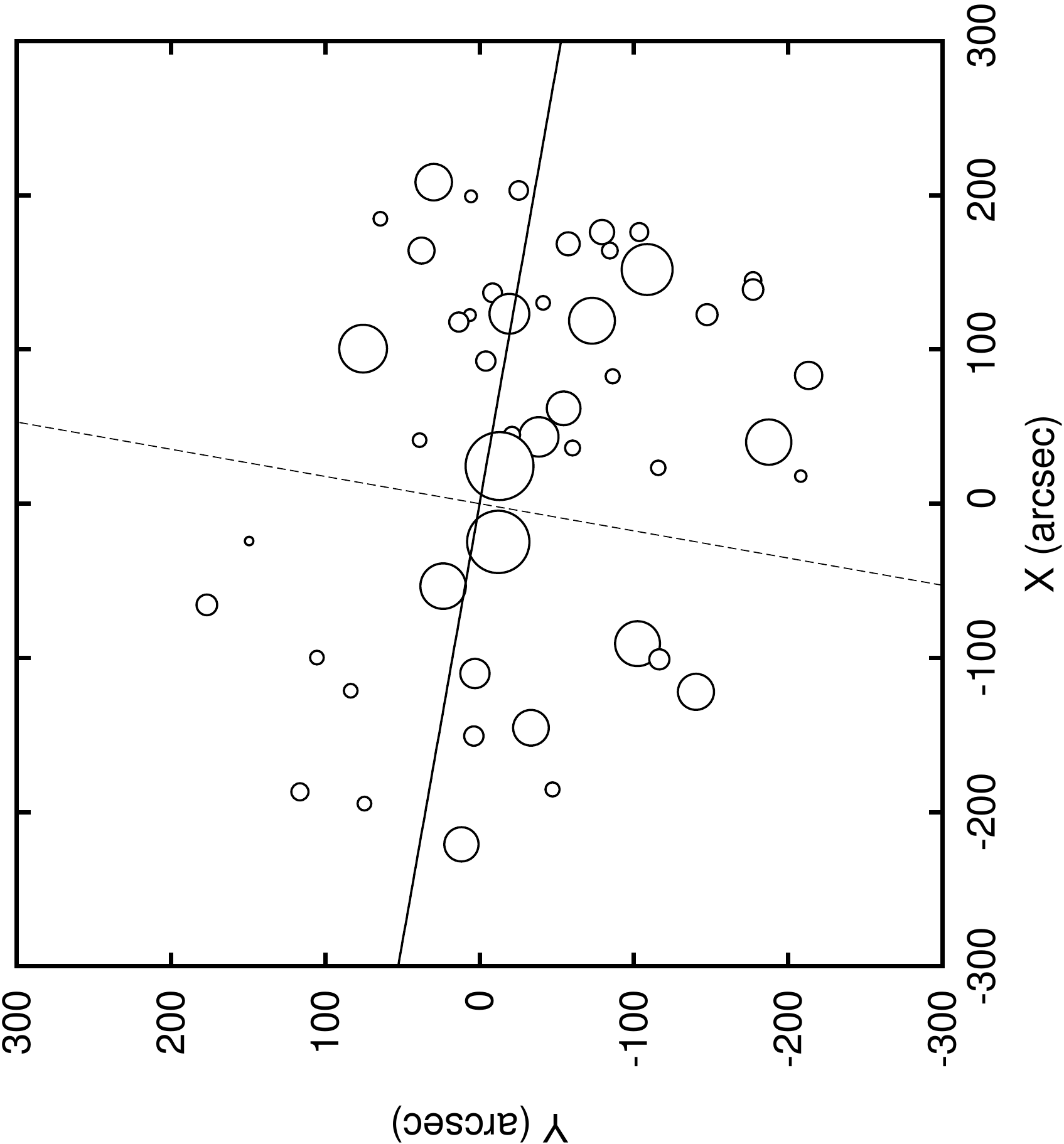}
    \caption{Projected spatial distribution of the GCs, where the RV magnitude is coded according to the size of the circles. Filled circles represents GCs with RV~\textgreater~1760\,km/s (left panel) and open circles represents GCs with RV~\textless~1760\,km/s (right panel). The continuous straight line indicates the axis of rotation of the GCs, perpendicular to the PA obtained with equation \ref{ecuacion1} (dashed line). The centre of the galaxy is at the origin of the coordinate system. North is up and East to the left.}
    \label{fig10n}
\end{figure*}
\begin{figure}
	\includegraphics[width=0.9\columnwidth,angle=270]{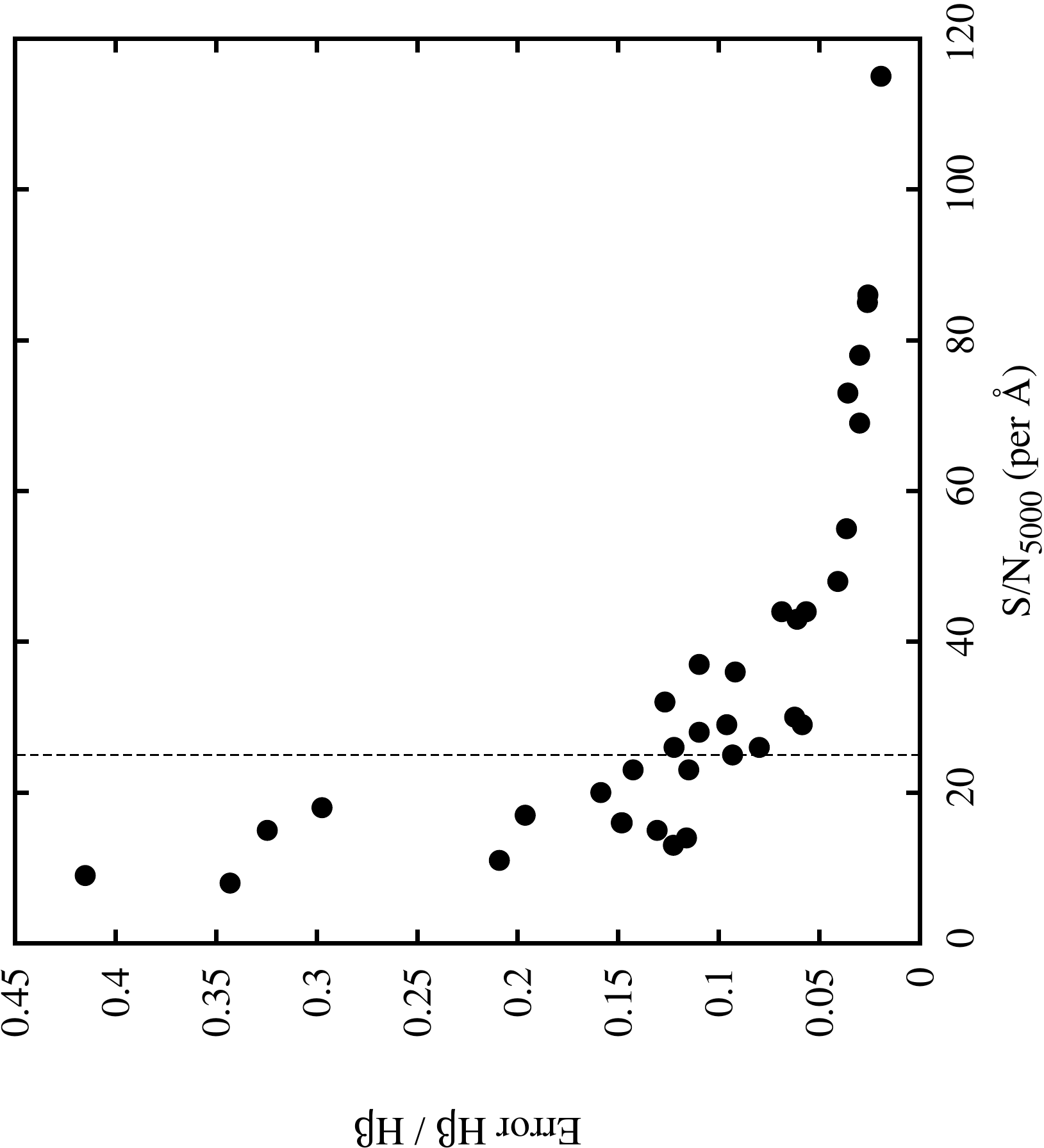}
    \caption{Relative error of the H$_\beta$ index as a function of the S/N per\,\AA\,(at 5000\,\AA) of each GC spectra. The dashed straight line at S/N per\,\AA~=~25 indicates the level of S/N adopted as a suitable value to obtain spectroscopic parameters with enough precision.}
    \label{fig11n}
\end{figure}

\subsection{RV distribution}
\label{sec3.3}

Figure \ref{fig7n} show the $(g-i)'_0$ colour distribution as a function of the RV for GCs of our sample. In this figure the GCs belonging to the subpopulation with \textit{red} and \textit{intermediate} colours seem to show a greater velocity dispersion than the subpopulation of \textit{blue} GCs, with respect to the systemic RV of NGC\,1316 (e.g. $\sigma_{interm}$~$\sim$~1.5$\sigma_{blue}$).
In order to analyze if this trend is significant, we performed a Kolmogorov-Smirnov (K-S) test to compare the RV distribution of the different GC subpopulations. In all cases, the hypothesis that the three samples are extracted from the same distribution cannot be rejected. However, it must be considered that this result may be influenced by the low number of objects in each group.

Figure \ref{fig8n} shows the radial velocity distribution as a function of the galactocentric radii for our data and the samples of \citet{Goudfrooij2001a} and \citet{Richtler2014}. In the case of the objects that were measured by different authors we opted, in the first place, for those obtained in this work, and secondly by those obtained by \citet{Goudfrooij2001a}. In this way, a representative sample was obtained for galactocentric radii from a few arcsec, to about 800 arcsec. The complete sample does not show structures that could give evidence of the presence of different subgroups. \citet{Goudfrooij2001a} pointed out the deviation from the systemic RV of those GCs with RV$\sim$1300\,km/s and R$_{gal}$\,\textless\,35\,arcsec (see Section 5.2.2 in \citealt{Goudfrooij2001a}), indicating that it could be due to a differential rotation in the inner regions, or simply due to a significant amount of radial motion for those innermost objects. The presence in our sample of the two GCs marked with the dotted ellipse in Figure \ref{fig8n}, would suggest to us that this last scenario would be the most probable. A third scenario would be that it is only due to a projection effect.

We quantify the rotation of the GC system making an analysis of the relation between the positional angle (PA) of the GCs and their RVs.
This procedure consists of fitting the rotation signal by the function
\begin{eqnarray}
\centering
RV = V_{sys} + v_0 \cdot cos(\theta - \theta_0)
\label{ecuacion1}
\end{eqnarray}

\noindent where V$_{sys}$ is the systemic velocity of the GC system, \textit{v$_0$}~is the rotation velocity amplitude, $\theta$ is the PA of each GC, and $\theta_0$ is the PA (measured from the North towards East) of the kinematic major axis of the GC system. This approach determines the best fit of a flat rotation curve as described by \citet{Zepf2000} and \citet{Cote2001}.

 In this instance the sample includes our data together with those GCs only presented by \citet{Goudfrooij2001a} and \citet{Richtler2014}. In order to make this sample as homogeneous as possible, only those GCs with R$_{gal}$~\textless~4\,arcmin in the \citet{Richtler2014} sample were included (apparent limit of the galaxy spheroid, \citealt{Schweizer1980}). Therefore the sample is composed of 35 objects from our data, 8 from \citet{Goudfrooij2001a} and 47 from \citet{Richtler2014}.
As a result we detect a rotation signal, whose set of parameters that provide the best-fit representation are shown in  Table \ref{table1b}.

%
\begin{table*}
\centering
\footnotesize
\begin{tabular}{lcc|cc|cl}
\hline
\hline
\multicolumn{3}{c}{GCs} &
\multicolumn{3}{c}{Stellar component} &
\multicolumn{1}{l}{}\\
\hline
V$_{sis}$[km/s]&v$_0$[km/s]&$\theta_0$[degrees]&$V_{gal}$[km/s]&$PA_{gal}$[degrees]&$\sigma$[km/s]&Reference\\
\hline
1735$\pm$17&145$\pm$32&$-$10$\pm$8&---&---&---&This work. Complete sample\\
1759$\pm$50&205$\pm$63&$-$9$\pm$3&---&---&---&This work. Intermediate colours\\
1698$\pm$48&179$\pm$68&1$\pm$17&---&---&---&\citet{Goudfrooij2001a}\\
1760*&120$\pm$64&$-$1$\pm$64&1760*&72&128&\citet{Richtler2014}\\
---&---&---&1760*&64$\pm$8&85$\pm$11&\citet{McNeil2012}\\
\hline
\end{tabular}
\caption{Kinematic fitting of globular cluster and stellar component velocities.(*)Previously fixed parameter.}
\label{table1b}
\end{table*}

%

This analysis was repeated considering only GCs that present GMOS photometry and that belong to the subpopulation with intermediate colours, i.e., 0.9~\textless~$(g-~i)'_0$~\textless~1.05\,mag (Figure \ref{fig9n}, right panel). Taking into account these criteria, the sample is composed of 17 GMOS, 11 VLT and 3 NTT GCs. Table \ref{table1b} shows a result similar to that obtained using the complete sample. This could be explained by the fact that the most of the objects to which the RV have been measured correspond to the brightest objects of the three samples, which have intermediate $(g-i)'_0$ colours, as shown in Figure \ref{fig3n}.

Perhaps Figure \ref{fig10n} is more illustrative to understand the rotation presented by the GCs. This figure shows the projected spatial distribution of the GCs, where the RV magnitude is coded according to the size of the circles. Objects with RV~\textless~1760\,km/s and RV~\textgreater~1760\,km/s are represented by open and filled circles, respectively. A tentative GC system rotation around an axis perpendicular to the PA obtained in Figure \ref{fig9n} can be seen in this diagram. Table \ref{table1b} summarizes the most relevant kinematic results present in the literature for both GCs and the stellar component of NGC\,1316. We will discuss the implication of these results in Section \ref{sec5.1}.

\section{Stellar populations}
\label{sec4}

\begin{figure*}
	\includegraphics[width=1.5\columnwidth]{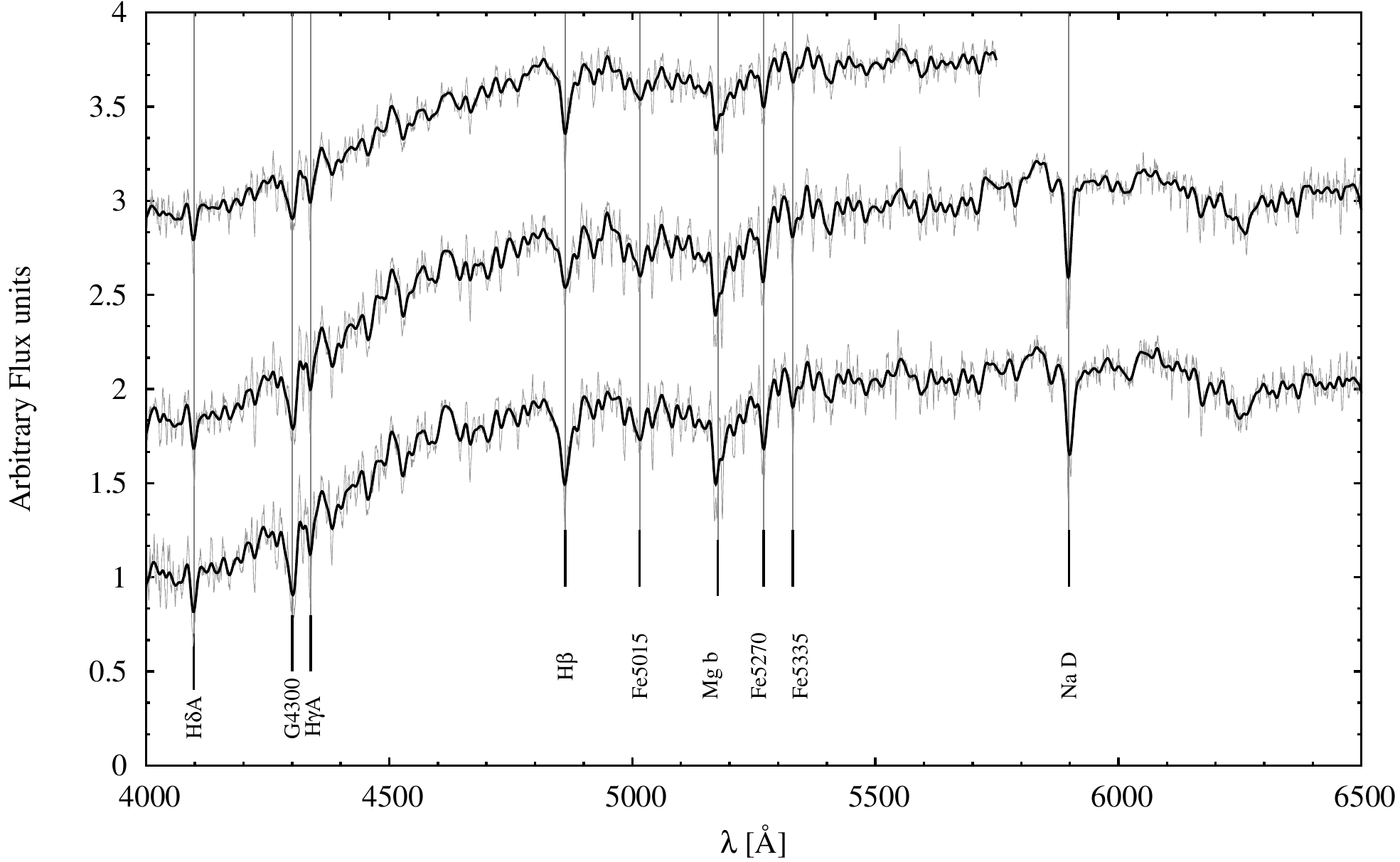}
    \caption{Normalized spectra, shifted arbitrarily in ordinates, for GCs ID$_{S}$= 1033, 156, 728 (from top to bottom) are shown in grey and the normalized spectra smoothed using a Gaussian kernel with an 4.5\AA~dispersion are shown in black. Vertical lines show the central wavelength of Lick/IDS indices H$\delta$A, G4300, H$\gamma$A, H$\beta$, Fe5015, Mgb, Fe5270, Fe5335 and Na D.}
    \label{fig12n}
\end{figure*}

In this section, we describe the measurement of ages, metallicities and $\alpha$-element abundances. The integrated properties of the GCs were determined by measuring the Lick/IDS absortion line indices \citep{Worthey1994, Worthey1997, Trager1998} for each spectrum and their subsequent comparison with SSP models. In this approach, we used the SSP models of \citet{Thomas2003,Thomas2004}, which have a spectral coverage between 4000$-$6500~\AA, ages from 1 to 15 Gyr and metallicities of [Z/H] = -2.25 to 0.67\,dex. One of the most outstanding features of these models is the fact that they include the effects produced by the abundance relations of $\alpha$-element. These models consider [$\alpha$/Fe] = 0.0, 0.3, 0.5\,dex. To carry out this task, we previously had to convolve our spectra with a wavelength-dependent Gaussian kernel to reproduce the Lick/IDS resolution \citep{Norris2008}. We measured the 25 Lick/IDS indices following the guidelines described in \citet{Worthey1997} and \citet{Trager1998} and subsequently we apply the offsets required to transform them to the Lick system. These values were obtained from the work of \citet{Loubser2009}. As an example, Figure \ref{fig12n} presents three spectra with high S/N per \AA, in which the central wavelength of some Lick/IDS indices are shown.

We adopted a minimum S/N per \AA~of 25 as a suitable value to obtain spectroscopic parameters with enough precision. This value is similar to that published in other works \citep[e.g.][]{Proctor2004,Pierce2006b,Beasley2008}. Figure \ref{fig11n} shows the relative error of the H$_\beta$ index as a function of the S/N per \AA. Final spectra with S/N per \AA~\textgreater~25 give relative errors in the H$_\beta$ index less than 15 percent. A similar behaviour is observed for the other indices.


\subsection{Diagnostic plots}
\label{sec4.1}

Diagnostic plots were used in order to obtain estimates of the age, metallicity and $\alpha$-element abundances of stellar populations present in NGC\,1316. These diagrams are constructed by selecting some Lick/IDS indices (or a set of them), which are overlaid with  \citet{Thomas2003,Thomas2004} synthetic SSP models. We build an  Mgb vs. \textless Fe\textgreater~ diagram, where \textless Fe\textgreater = [(Fe5270 + Fe5335)/2] \citep{Gonzalez1993}. We use this set of indices since this combination is insensitive to age and metallicity, but is very useful in limiting the $\alpha$-element abundances \citep{Kuntschner2002}. Furthermore, we constructed an [MgFe]' vs. H$\beta$ diagram, where [MgFe]'=$\sqrt{Mgb\, (0\ldotp72\cdot Fe5270 + 0\ldotp28\cdot Fe5335)}$~ \citep{Thomas2003} which stands out for being completely independent of the value of [$\alpha$/Fe].
%
\begin{figure}
	\includegraphics[width=0.9\columnwidth,angle=270]{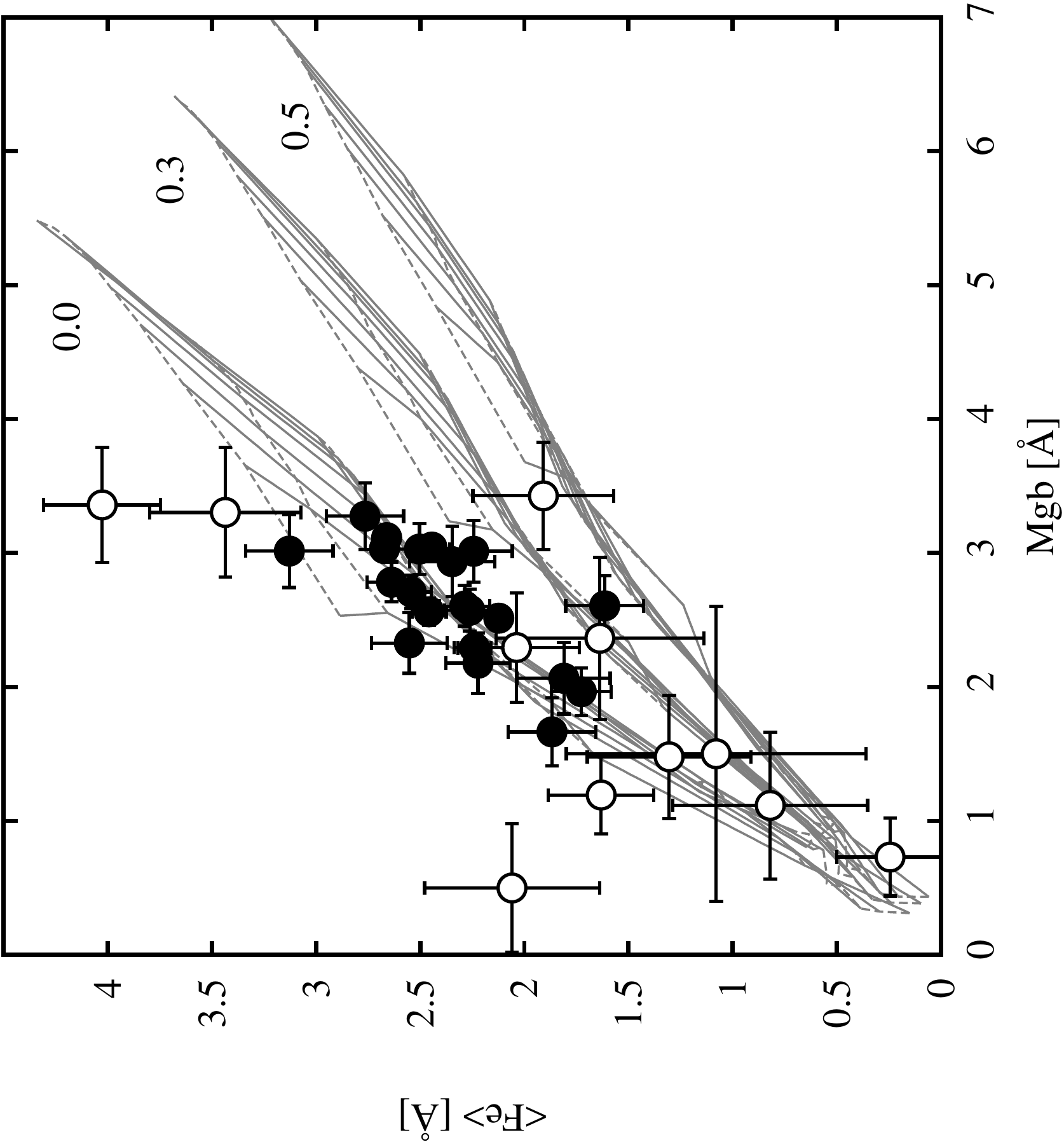}
    \caption{Diagnostic plot Mgb vs. \textless Fe\textgreater~ for the 35 confirmed GCs (see Section \ref{sec3.1}). We distinguish between GCs with spectra with S/N per\,\AA~\textless~25 (open circles) and S/N per\,\AA~\textgreater~25 (filled circles). Overplotted are models by \citet{Thomas2003, Thomas2004} with abundance ratios [$\alpha$/Fe]= 0.0, +0.3, +0.5, ages of 1, 3, 5, 8 and 12 Gyr and metallicity [Z/H]= -2.25, -1.35, -0.33, 0.0, +0.35  and +0.67\,dex.}
    \label{fig13n}
\end{figure}
\begin{figure*}
	\includegraphics[width=\columnwidth,angle=270]{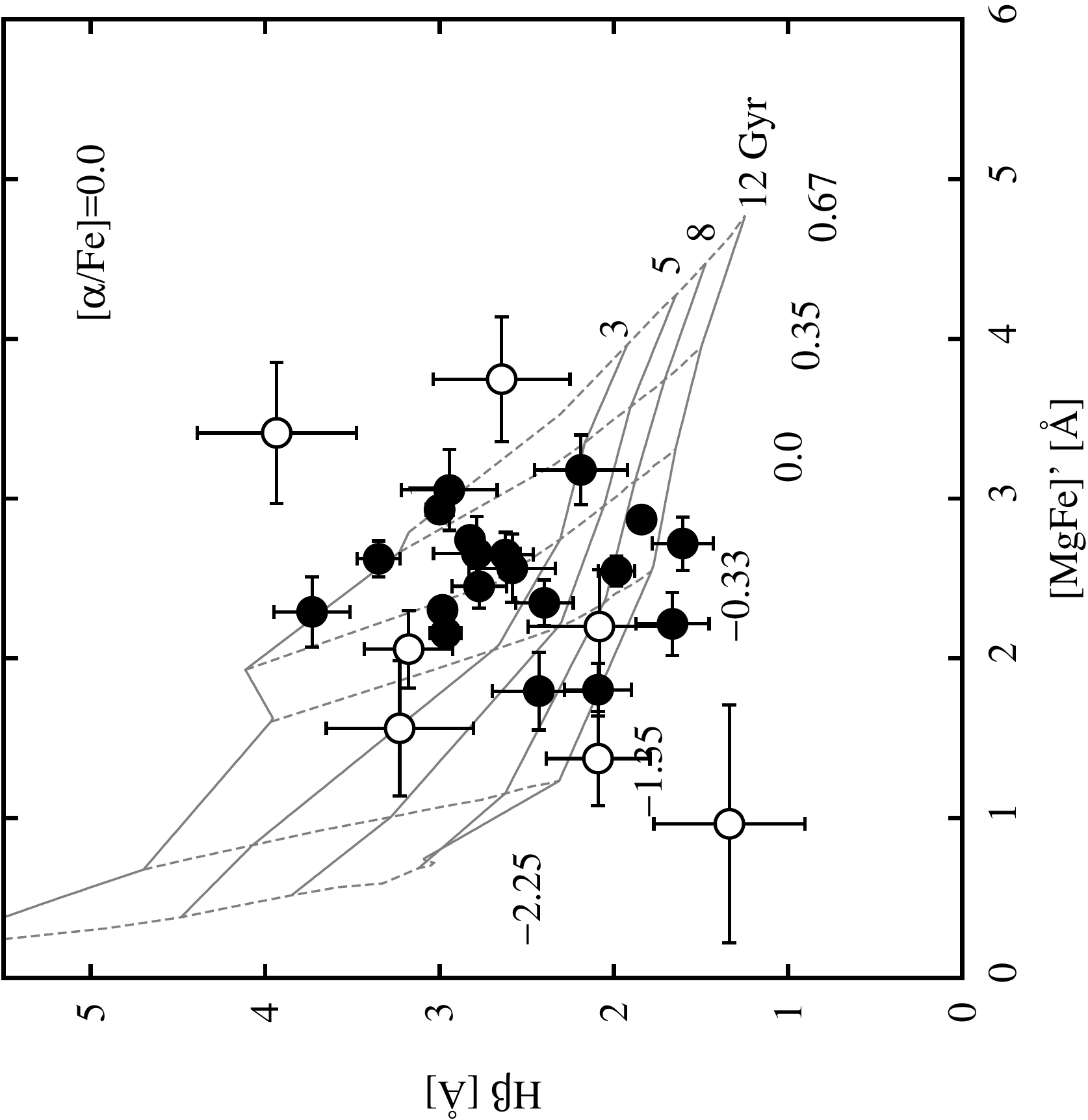}
	\includegraphics[width=\columnwidth,angle=270]{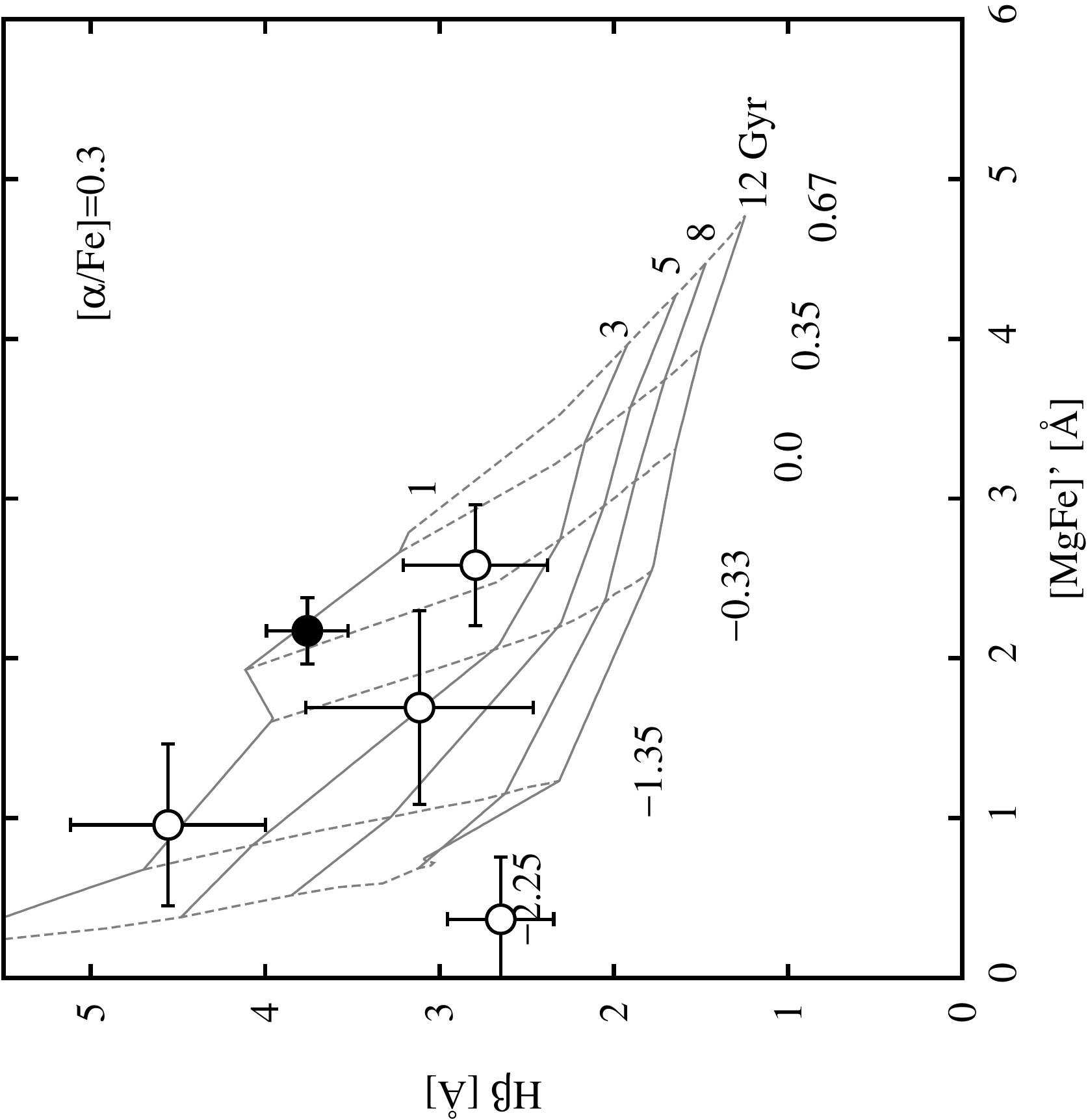}
    \caption{Diagnostic plots [MgFe]' vs. H$\beta$ for those GCs with [$\alpha$/Fe]= 0.0 (\textit{left panel}) and [$\alpha$/Fe]= 0.3 (\textit{right panel}). We distinguish between GCs with spectra with S/N per\,\AA~\textless~25 (open circles) and S/N per\,\AA~\textgreater~25 (filled circles). Overplotted are models by \citet{Thomas2003, Thomas2004} with fixed abundance ratios [$\alpha$/Fe], ages of 1, 3, 5, 8 and 12 Gyr and metallicity [Z/H]= -2.25, -1.35, -0.33, 0.0, +0.35  and +0.67\,dex.}
    \label{fig14n}
\end{figure*}

Figure \ref{fig13n} presents the diagnostic plot  Mgb vs. \textless Fe\textgreater \, for the 35 GCs confirmed in the previous section. Most of them appear to agree with a value of [$\alpha$/Fe]= 0.0, and a small group has a value of [$\alpha$/Fe]= 0.3. The difference becomes more significant if one takes into account only those objects with high S/N.

Meanwhile, Figure \ref{fig14n} shows [MgFe]' vs. H$\beta$ for those GCs that have a value of [$\alpha$/Fe]= 0.0 (left panel) and [$\alpha$/Fe]= 0.3 (right panel). In the first case two groups can be observed, one composed of young objects between 1 and 3 Gyr and metallicities between -0.3 and 0.67 dex, and another one of older objects with ages between 5 and 12 Gyr and metallicities between -1.35 and 0.0 dex. In contrast, the evident spread presented  by those GCs with  [$\alpha$/Fe]= 0.3 is due to the fact that most of its spectra have low S/N. Only one GC presents S/N per\,\AA~\textgreater~25 (ID$_{Sesto}$~=~94), which would be a young object with an age of $\sim$ 1 Gyr and solar metallicity.

\subsection{Ages, metallicities and $\alpha$-element abundances}
\label{sec4.2}

In order to determine the age, metallicity and $\alpha$-element abundances of each GC, the $\chi^2$ minimization method of \citet{Proctor2002} and \citet{Proctor2004} was used. This technique simultaneously compares the different observed Lick/IDS indices with those obtained from SSP models, selecting the combination that minimizes the residuals through a $\chi^2$ fitting. In this particular case, we used the SSP models of \citet{Thomas2003,Thomas2004} previously mentioned.

To estimate the integrated properties of each GC, we used those spectral indices that presented the smallest  relative errors. These were selected from the group conformed by H$\delta$A, H$\delta$F, H$\gamma$A, G4300, Fe4383, H$\beta$, Fe5015, Mgb, Fe5270, Fe5335 and Fe5406 since they provide acceptable results. In all cases, a minimum of 5 indices were used. The values of age, metallicity and chemical abundance are presented in Table \ref{table3}. Their uncertainties were estimated using 100 Monte Carlo re-simulations, which include estimates of the errors in the indices according to the obtained S/N per\,\AA.

The spectroscopic results were combined with the photometric data obtained in Paper I, in order to determine the characteristics of the different GC subpopulations.

\subsubsection{Reddening of the innermost GC}
\label{sec4.2.0}

\begin{figure}
\centering
	\includegraphics[width=0.99\columnwidth]{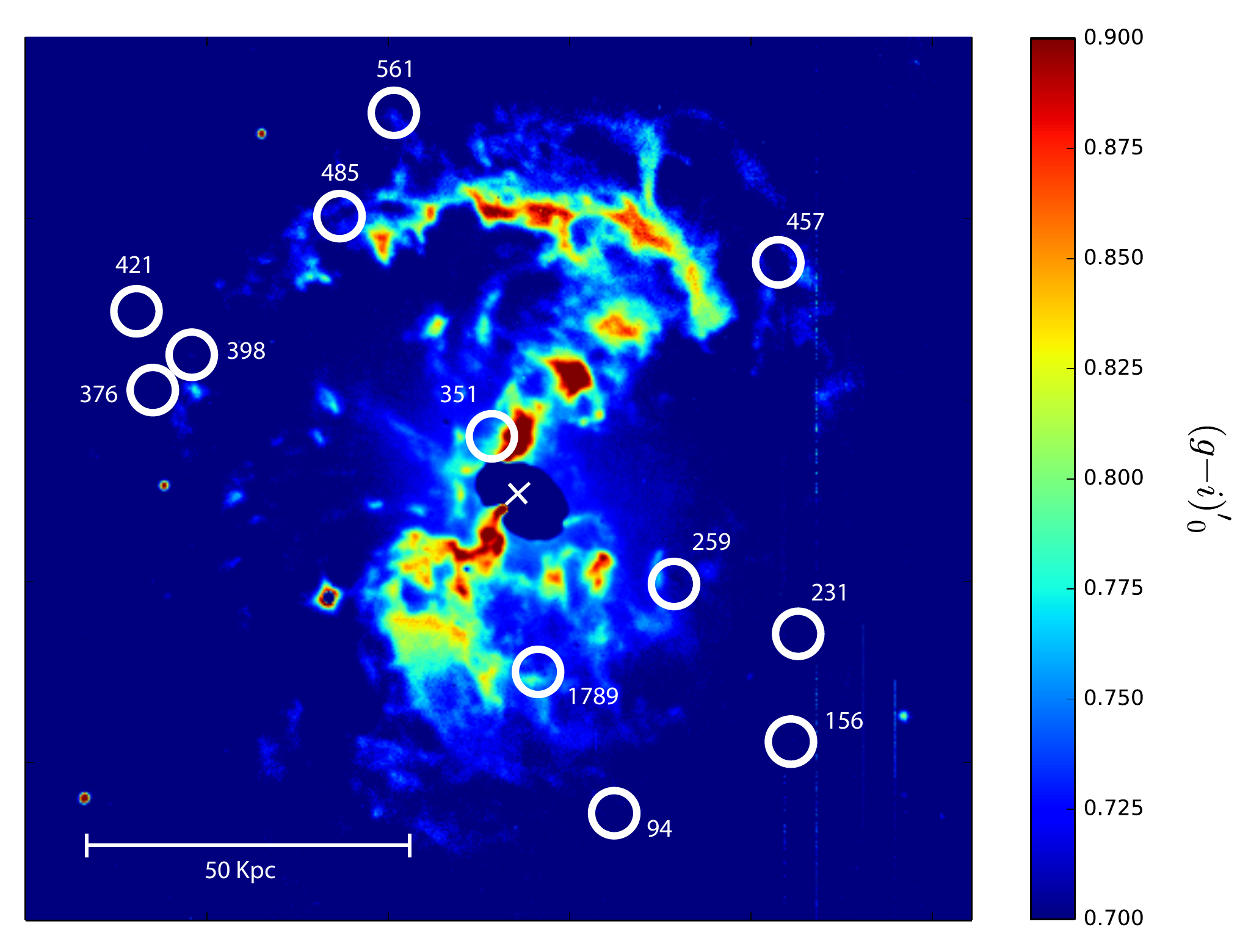}
    \caption{Colour map of the central region of NGC\,1316. The colorbar indicate the  $(g-i)'_0$ colour. The centre of NGC\,1316 is displayed with a cross. North is up and East to the left. White circles shows the positions of the innermost GCs, with their respective labels (see Table \ref{table3}).
}
    \label{fig15n}
\end{figure}

Figure \ref{fig15n} shows the colour map $(g-i)'_0$ of the central region of NGC\,1316. We can observe a complex distribution of the dust, with a particular structure that imitates a bar with pronounced spiral arcs \citep{Schweizer1980, Iodice2017}. The figure shows the positions of the GCs closest to this structure (white circles) with their respective identification.

Analyzing the colour map and the position of the objects in Figure \ref{fig3n}, the GC ID$_S$=351 is the only one that shows evidence of being significantly affected by dust. 
Taking into account that dust extinction effects are small for most Lick/IDS indices \citep{Macarthur2005}, we estimate the apparent reddening of this object calculating its synthetic colours through SSP models. Although there are different SSP models available in the literature, in \citet{Sesto2016} it was determined that the PARSEC models \citep{Bressan12} are in good agreement with our $gri'$ photometry. In turn, in that work we discussed the need to make a small correction to the PARSEC colour indices including the $g'$-band magnitudes.
The comparison between the synthetic and the observed $(g-i)'$ colour indicates that this GC seems to be reddened by $\sim$0.5\,mag.

\begin{figure}
	\includegraphics[width=\columnwidth]{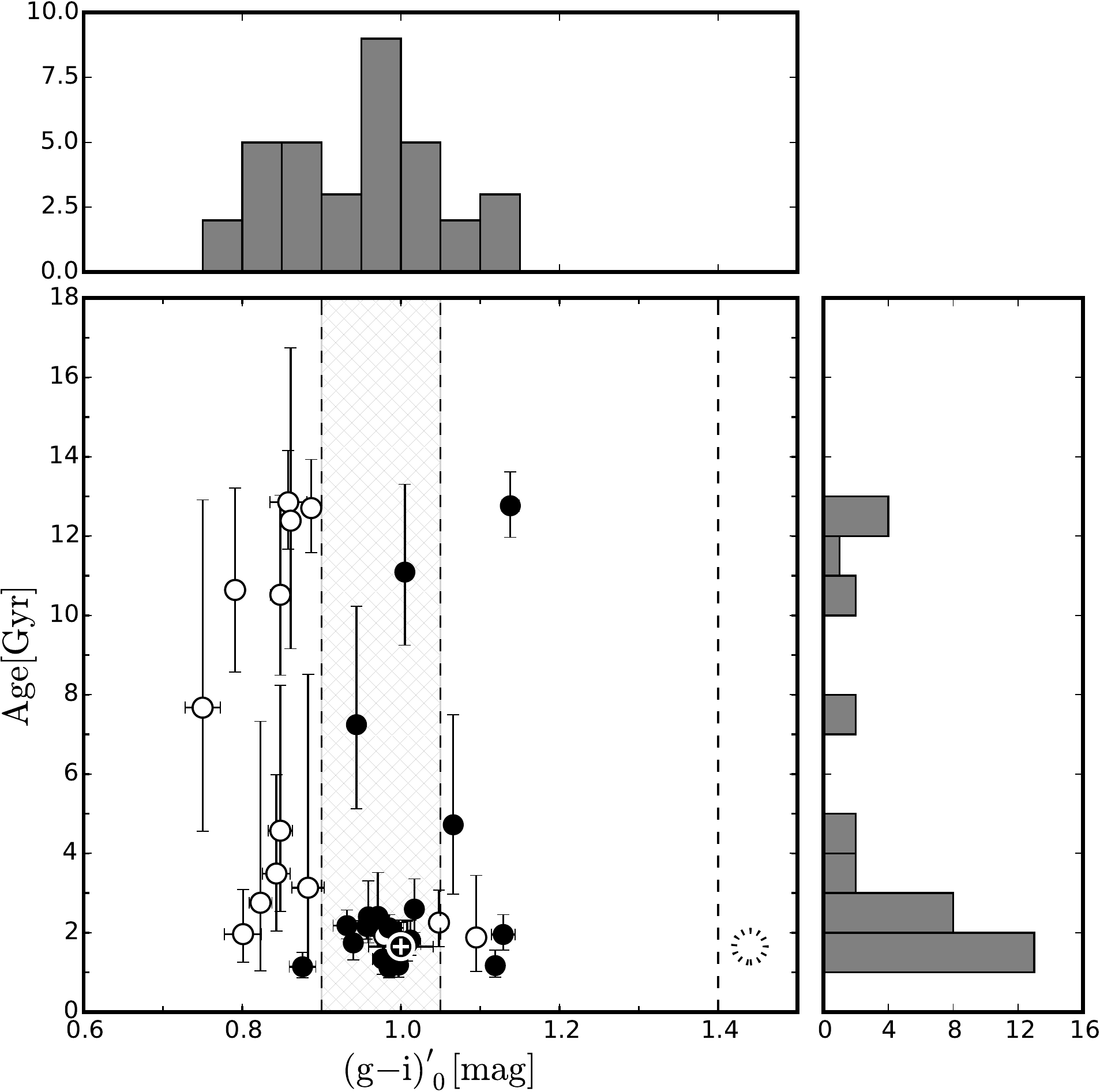}
 
\caption{Ages of the GCs as a function of colour $(g-i)'_0$. We distinguish between GCs with spectra with S/N per\,\AA~\textless~25 (open circles) and S/N per\,\AA~\textgreater~25 (filled circles). The object ID$_S$=351 is shown with colour uncorrected by reddening (dashed circle) and corrected following the guidelines of Section \ref{sec4.2.0} (white plus sign). The scratched area indicate the colour ranges for the intermediate-colours GCs. The dashed line shows the limit established in paper I for the \textit{red} GCs. The boxes above and to the right show the histograms of colour and age for all objects.}
    \label{fig16n}
\end{figure}

\begin{figure*}
	\includegraphics[width=0.85\columnwidth, angle=270]{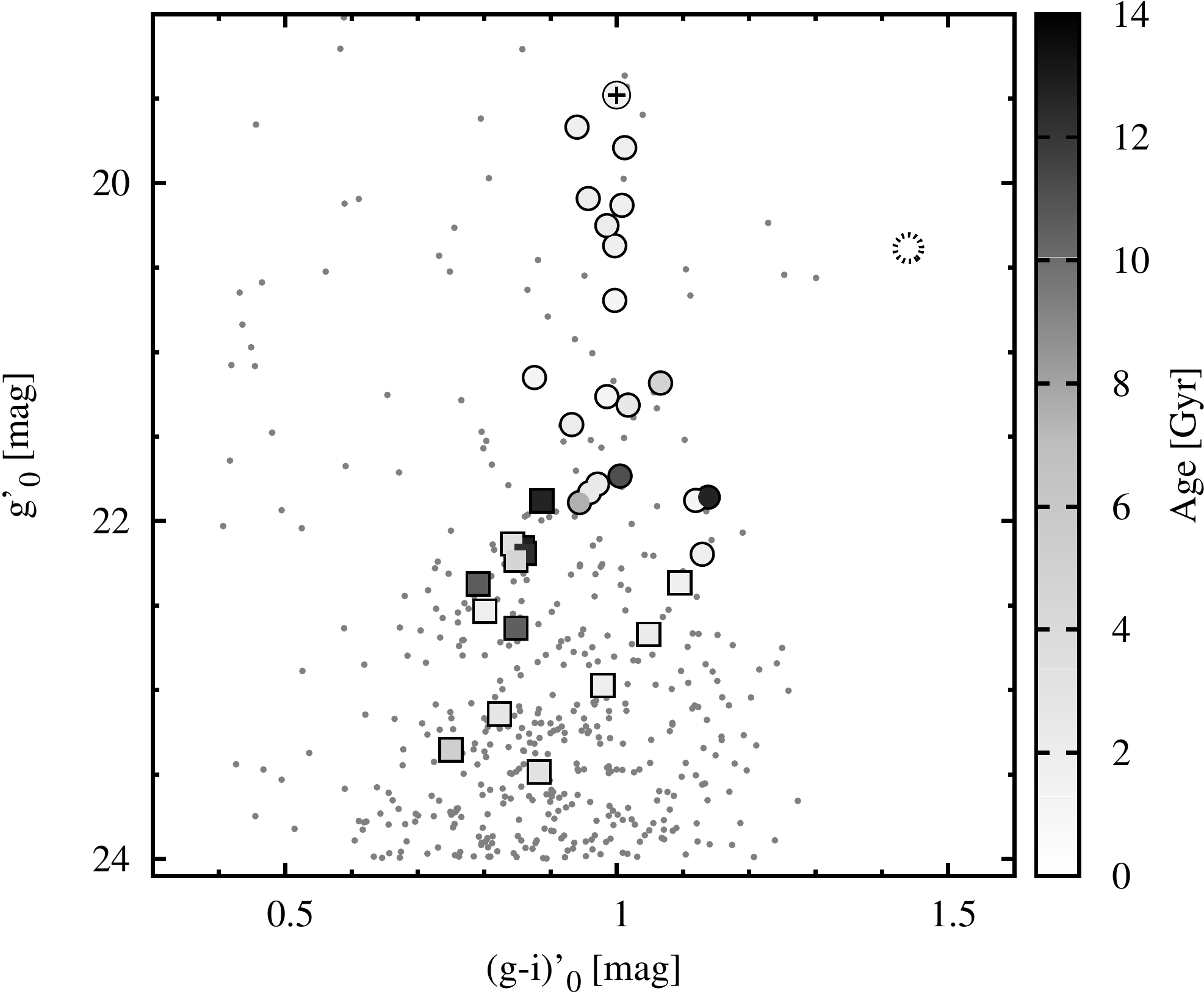}
	~~~\includegraphics[width=0.85\columnwidth, angle=270]{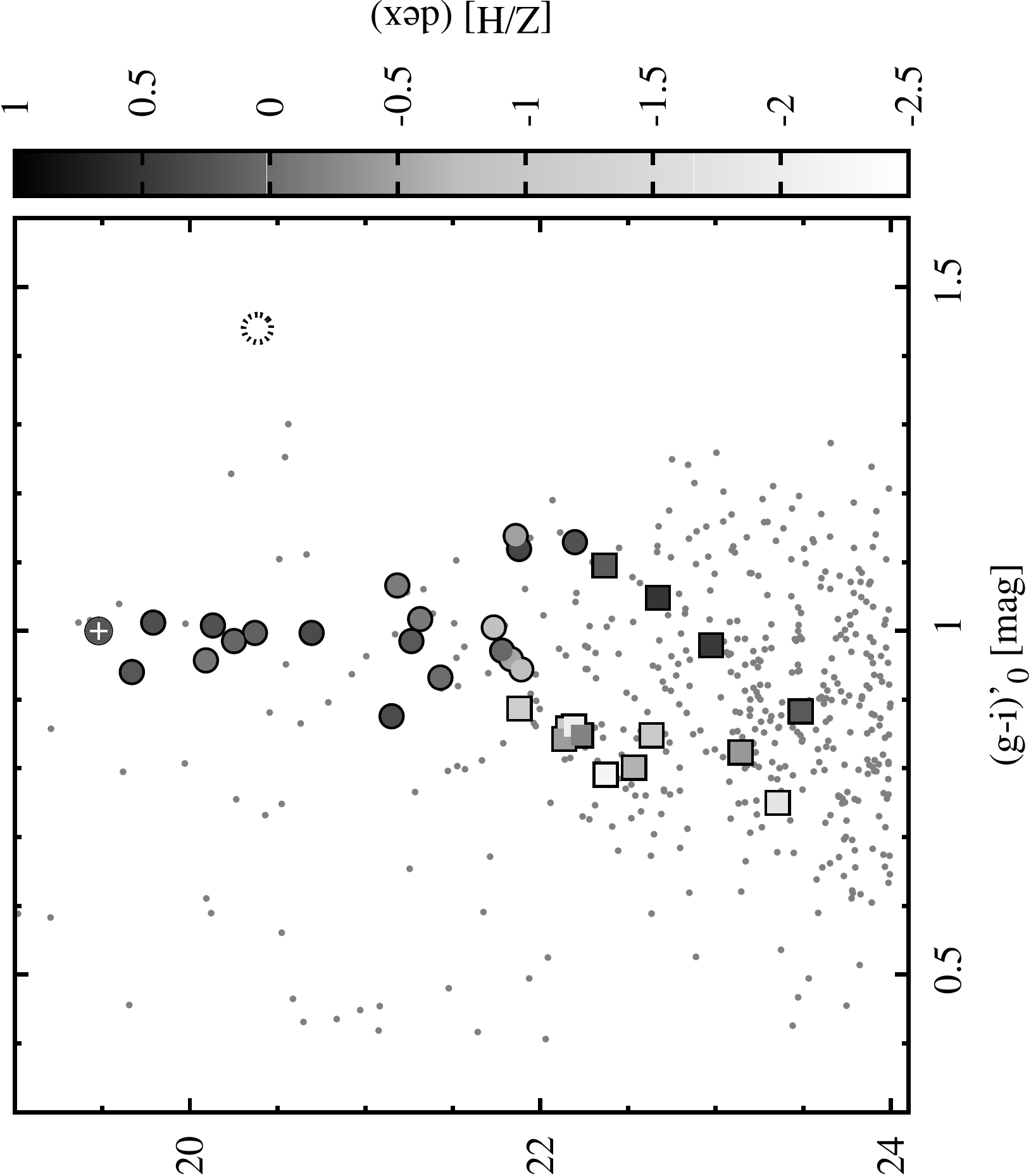}
 \caption{CMD of the photometric and spectroscopic data. The small grey circles correspond to the GCs candidates brighter than $g'_0$ = 24\,mag (paper I). The greyscale indicate the different ages (left panel) and metallicities (right panel) of the GCs with spectroscopic information, as is indicated in the sidebar of the figure. The object ID$_S$=351 is shown with colour uncorrected by reddening (dashed circle) and corrected following the guidelines of Section \ref{sec4.2.0} (plus sign). We distinguish between GCs with spectra with S/N per\,\AA~\textless~25 (squares) and S/N per\,\AA~\textgreater~25 (circles).}
    \label{fig17n}
\end{figure*}

Taking into account R$_V=$~1.3, we determine the extinction coefficient corresponding to the filter $g'$ (A$_{g'}$) using the relations $\frac{A_{g'}}{E(B-V)}$~=~3.303 and $\frac{A_{i'}}{E(B-V)}$~=~1.698 \citep{Schlafly2011}. Thus, we obtained for this GC a tentative approximation of both, the $(g-i)'_0$ colour and the $g'$-filter magnitude in the absence of dust. Figures \ref{fig16n}, \ref{fig17n} and \ref{fig18n} show the position of this object with colour uncorrected (dashed circle) and corrected by reddening (white plus sign). The corrected position in the different diagrams is in agreement with the rest of the young GC sample. Another interesting result is that, as seen in Figure \ref{fig17n}, ID$_S$=351 turns out to be the brightest object in our spectroscopic sample.


\subsubsection{Ages}
\label{sec4.2.1}

Figure \ref{fig16n} shows the age of the GCs as a function of colour $(g-i)'_0$. Dashed lines indicate,  taking into account what was obtained in Paper I, the colour ranges with which the different subpopulations of GCs were separated (see Section \ref{Intro}). With these arguments in mind, the spectroscopic sample is composed of 12 \textit{blue}, 17 \textit{intermediate} and 6 \textit{red} GCs. Most of the sample belongs to the subpopulation of GCs with intermediate colours, being more obvious in those with S/N per\,\AA~\textgreater~25. The right panel histogram shows the presence of two age-differentiated groups, where a crowd of young GCs with ages between 1 and 6 Gyr stands out. This group amazingly represents 80$\%$ of the sample. GCs with ages between 6 and 13 Gyr make up the rest of the sample, although to a lesser extent. As mentioned in the introduction, it is possible to find young and intermediate-age GCs in the GC systems of some merger remnants \citep[e.g.][]{Schweizer2002, Strader2004, Woodley2010a}, but in all of them this population consists of only a few objects and they are far from reaching levels over the total of the sample as we have found in the GC system of NGC\,1316.

Figure \ref{fig17n} displays the CMD of the photometric and spectroscopic samples. Small grey dots represent unresolved sources brighter than $g'_0$ = 24\,mag. In the left panel the greyscale indicates the different ages of the GCs with spectroscopic information. The sample is dominated by young GCs, especially in those with high S/N, where the estimation of integrated properties has a higher level of reliability. In addition, the large number of GCs that are brighter than $\omega$centauri, the brightest GC of the MW, ($g'_0$~\textless~21.6\,mag) stands out. As mentioned in Section \ref{sec3.1}, this may be due to a size-of-sample effect, but also to the effect reported by \citet{Whitmore1997}. However, it is necessary to emphasize the remarkable narrow range in colour that the brighter GCs present. As will be seen in the next section, this is mainly due to the fact that these objects, in addition to being young, have solar metallicity or greater.

\subsubsection{Metallicities}
\label{sec4.2.2}

\begin{figure}
	\includegraphics[width=0.9\columnwidth, angle=270]{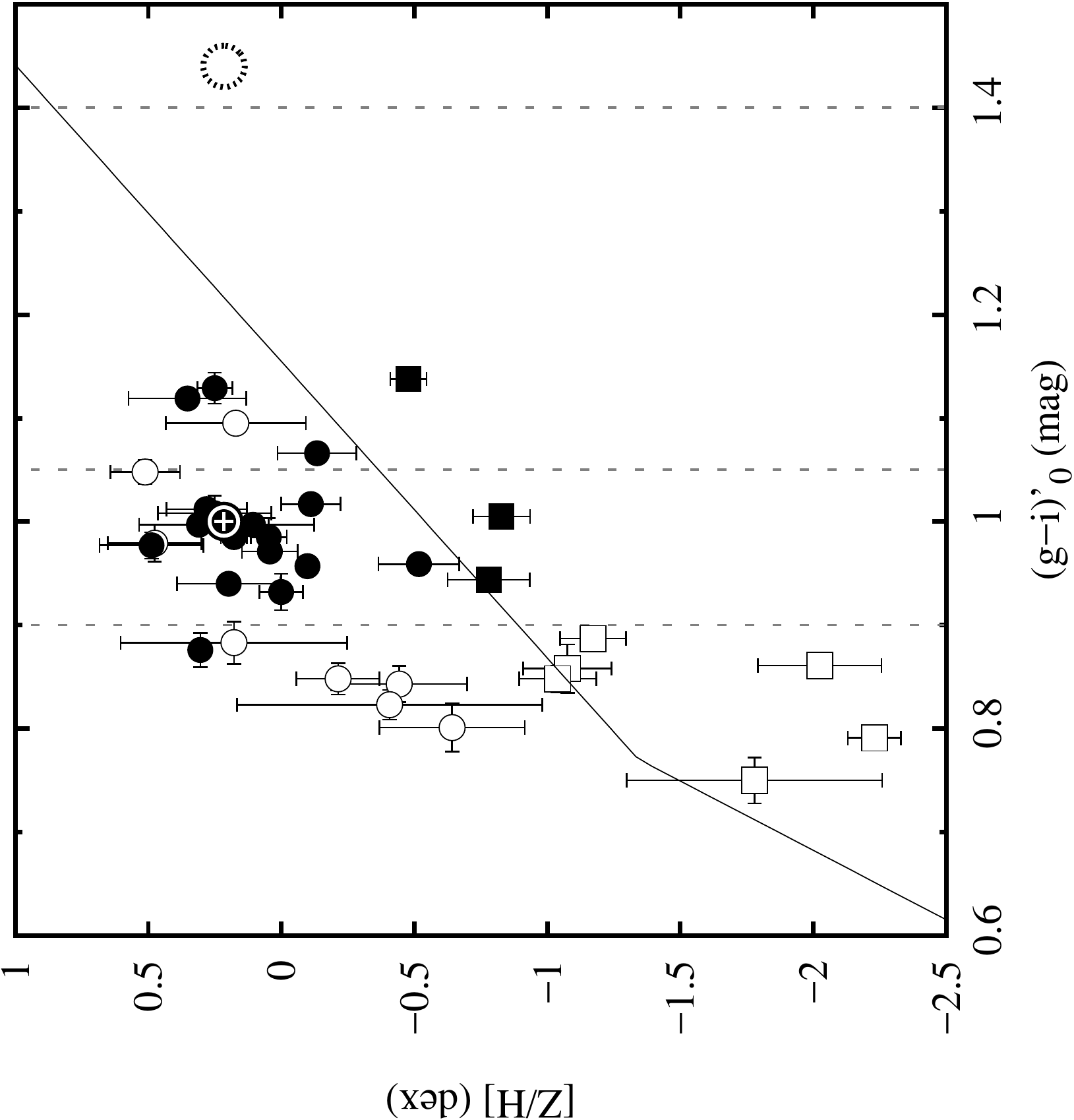}
 \caption{Metallicity of the GCs as a function of colour $(g-i)'_0$. We distinguish between GCs with spectra with S/N per\,\AA~\textless~25 (open symbols) and S/N per\,\AA~\textgreater~25 (filled symbols).
Circles represent objects with ages between 1 and 6 Gyr, and squares represent objects with ages between 6 and 13 Gyr. Dashed lines indicate the colour ranges of the different subpopulations of GCs (see text). The object ID$_S$=351 is shown with colour uncorrected by reddening (dashed circle) and corrected following the guidelines of Section \ref{sec4.2.0} (white plus sign).
The continuous line corresponds to the calibrations of \citet{Usher2012}.}
    \label{fig18n}
\end{figure}

We analyzed the metallicity of the GCs as a function of colour $(g-i)'_0$. Figure \ref{fig18n} shows this analysis, following the guidelines presented in Figure \ref{fig16n}. It is observed that the objects  in our sample present metallicities from -2 to +0.5 dex, covering the whole range of the used models. In the particular case of spectra with S/N per\,\AA~\textgreater~25, GCs presents relatively high metallicities, i.e. -1~\textless~[Z/H]~\textless~0.5\,dex. In turn the sample is differentiated between GCs with ages smaller and greater than 6 Gyr. The figure shows the [Z/H] vs. $(g-i)'_0$ colour relation  based on spectroscopic observations of the Calcium triplet lines presented by \citet{Usher2012} (continuous ``broken'' line). The GC sample used by these authors is presumably dominated by old (12 Gyr) clusters. As expected, the young GCs fall above this relation, since at equal metallicity the young GCs must be located on the blue side of the ``broken'' line. Instead, the older GCs seem to fall slightly below the empirical colour-metallicity relation. We cannot rule out the existence of a small difference of zero point between our photometry and that observed by \citet{Usher2012}. In that sense applying a small offset of 0.05\,mag towards the blue in $(g-i)'$, our sample of old GCs seems to be in a well agreement with the \citet{Usher2012} relation. At the red end, the dashed circle shows the location that would occupy the GC ID$_s$=351 if its colour is not corrected by reddening. This behaviour would not fit with any of the two subpopulations mentioned in this analysis.

The CMD of the photometric and spectroscopic samples is shown in the right panel of Figure \ref{fig17n}. An interesting result is the fact that the group of GCs that are brighter than $g'_0$= 21.6\,mag, which is composed mostly of young objects, has in all cases relatively high metallicities ([Z/H]~\textgreater~$-$0.5\,dex).

%
\begin{figure}
	\includegraphics[width=\columnwidth]{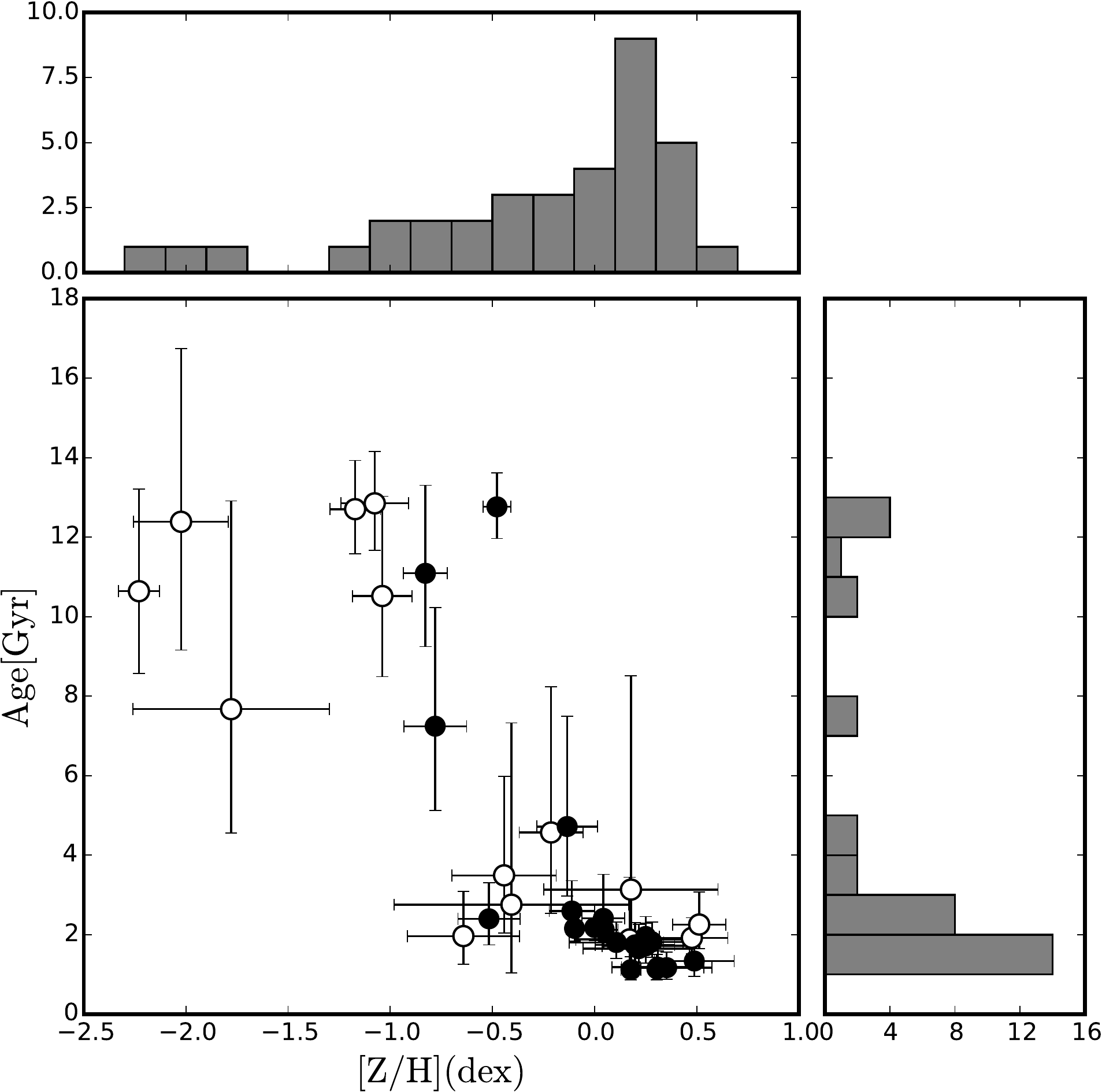}
 \caption{[Z/H] vs. age for objects with S/N per\,\AA~\textless~25 (open circles) and S/N per\,\AA~\textgreater~25 (filled circles). The histogram of metallicities in the upper box shows that most of the objects have metallicities between $-$0.5 and +0.5\,dex.}
    \label{fig19n}
\end{figure}

Figure \ref{fig19n} shows the relationship between age and metallicity, where the different characteristics of the two main subpopulations mentioned above are easily observable: a subpopulation of 1$-$6 Gyr, and metallicities between $-$1~\textless~[Z/H]~\textless~0.5 dex and a smaller group of older GCs with ages between 6$-$13 Gyr which exhibit a spread of metallicities from $-$2.5 to $-$0.45\,dex. 
In the latter group, the objects with spectra showing S/N per\,\AA~\textgreater~25 presents metallicities from $-$1 to $-$0.45\,dex where we observed the presence of a classic \textit{red} GC (ID$_s$=636), that is, with age greater than 10 Gyr and [Z/H]~\textgreater~$-$0.6\,dex.

\subsubsection{$\alpha$-element abundances}
\label{sec4.2.3}

The generation of $\alpha$-elements (O, Ne, Mg, Si, S, Ar, Ca and Ti) originates mainly in the collapse of massive stars (type II supernovae), whereas the contribution of heavier elements (Fe, Cr, Mn, Fe, Co, Ni, Cu and Zn) are derived from progenitors with longer lifetimes (type Ia supernovae). Therefore, the abundance ratio [$\alpha$/Fe] reflects the effect of two different mechanisms of nucleosynthesis, and gives us an idea of how fast the formation of these objects was.

\begin{figure}
	\includegraphics[width=\columnwidth]{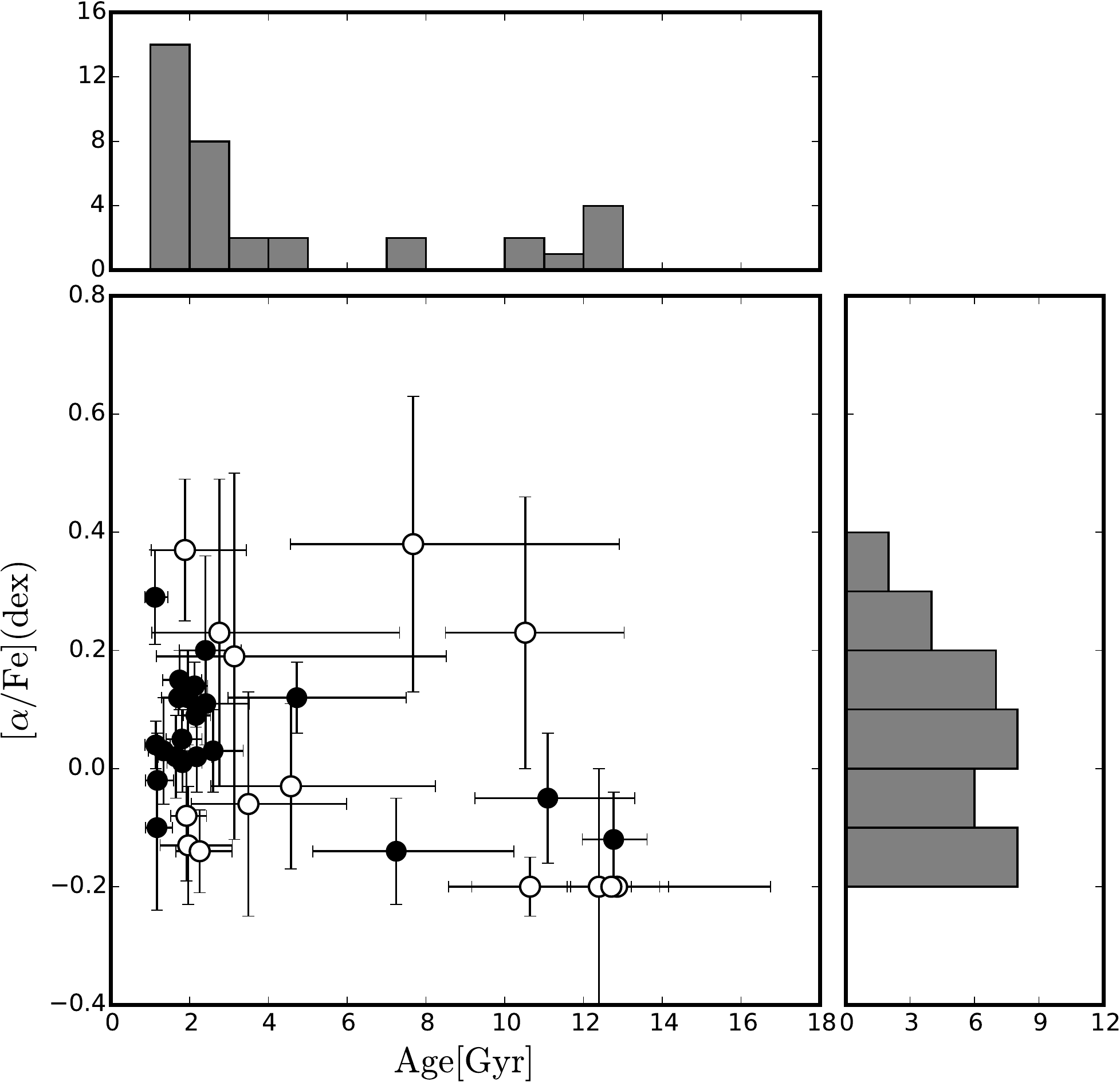}
 \caption{Age vs. [$\alpha$/Fe] for objects with S/N per\,\AA~\textless~25 (open circles) and S/N per\,\AA~\textgreater~25 (filled circles). The right box shows the histogram of $\alpha$-element abundancies.}
    \label{fig20n}
\end{figure}

\begin{figure}
	\includegraphics[width=\columnwidth]{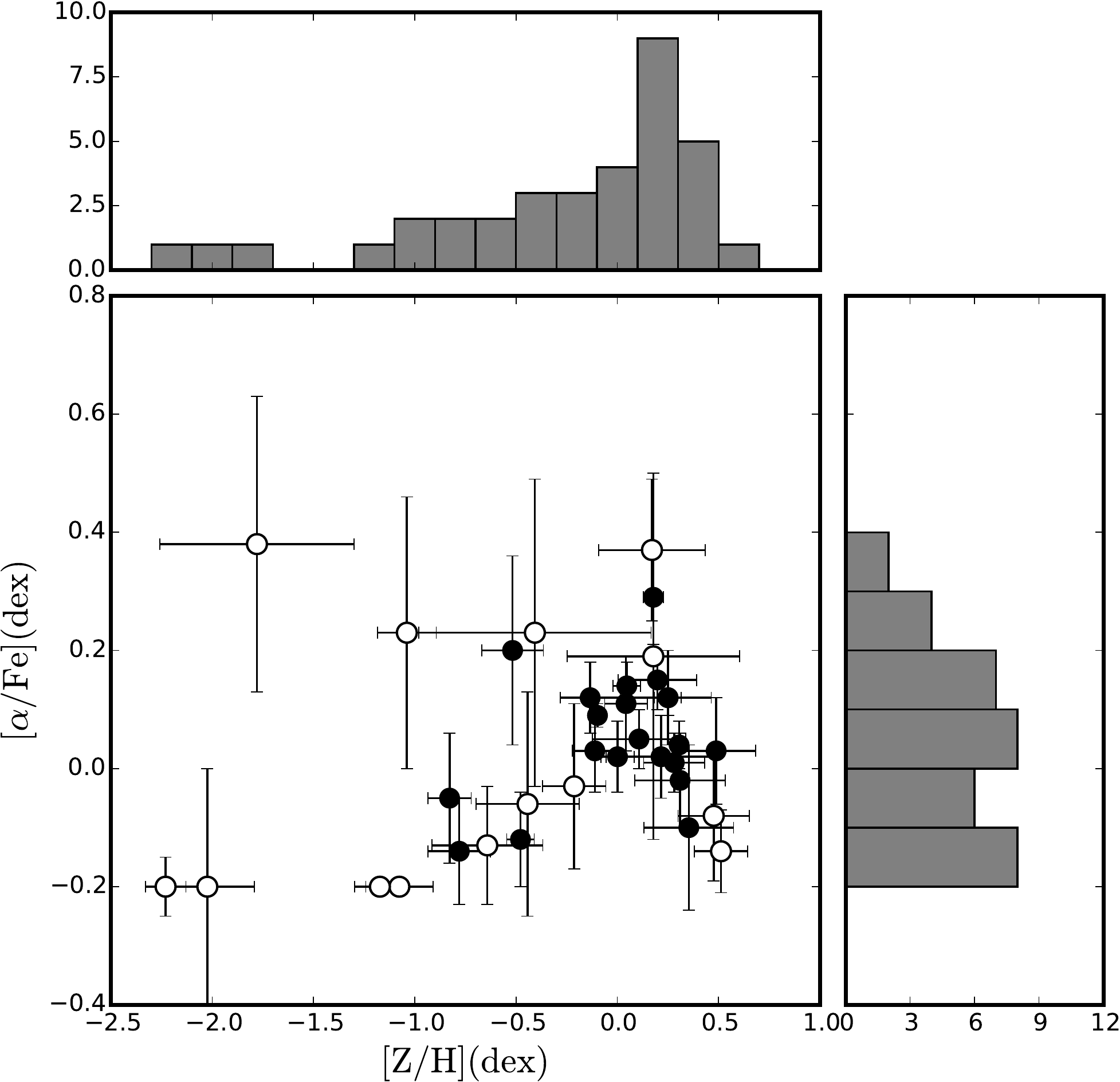}
 \caption{[Z/H] vs. [$\alpha$/Fe] for objects with S/N per\,\AA~\textless~25 (open circles) and S/N per\,\AA~\textgreater~25 (filled circles). It is relevant to note that as [Z/H] decreases, it is more difficult to obtain a good estimate of [$\alpha$/Fe].}
    \label{fig21n}
\end{figure}

We have analysed the variation of [$\alpha$/Fe] with age and metallicity. Figure \ref{fig20n} and \ref{fig21n} shows that young GCs have values of [$\alpha$/Fe] from -0.2 to 0.3 dex. Looking in detail we could distinguish two possible groups: one with [$\alpha$/Fe]\,$\sim$~0.1, and another group just above zero. However, due to the low number of objects, and the small separation between these possible subgroups, the results of several statistical tests (GMM; \citealt{Muratov2010} and RMIX\footnote{RMIX is publicly available at http://www.math.mcmaster.ca\\
/peter/mix/mix.html}) were not conclusive.


%
%

In turn those GCs with ages older than 6 Gyr seem to presents metallicity values in the range -2~\textless~[Z/H]~\textless~0.0\,dex (as mentioned above) and [$\alpha$/Fe] from $-$0.2 to 0.1\,dex, although the errors are significant. These results are consistent with that observed in diagnostic plots of Section \ref{sec4.1}.

Unlike what was published for the galaxies NGC\,3379 \citep{Pierce2006a}, NGC\,4649 \citep{Pierce2006b} and the sample of GCs in early-type galaxies by \citet{Puzia2005b}, we did not observe a trend between metallicity and $\alpha$-element abundances. The lack of this relation is also observed in other GC samples, as in the case of NGC\,1407 \citep{Cenarro2007}, NGC\,3923 \citep{Norris2008} and NGC\,5128 \citep{Woodley2010a}.

 There is a group of very interesting objects, which presents [Z/H]\,\textless\,$-$0.3 and subsolar values of [$\alpha$/Fe]. In Figure \ref{fig20n} these objects show a spread in their age (ranging from a few Gyr to almost 13 Gyr). 
As it will be discussed in Section \ref{sec5.2.1}, these characteristics have been observed in nuclei of dwarf elliptical galaxies (e.g. \citealt{Paudel2011}).


\section{Discussion}
\label{sec5}

\subsection{GCS Rotation}
\label{sec5.1}

The previous works of \citet{Goudfrooij2001a} and \citet{Richtler2014} performed studies of the rotation of the NGC\,1316 GC system. The first analyzed the rotation of the innermost component of the GC system (R$_{gal}$~\textless~25 Kpc). The latter performed this same analysis at different galactocentric radii, but only found evidences of rotation for those GCs with R$_{gal}$~\textless~25 Kpc and R-filter~\textgreater~22\,mag. In this work, new estimates of rotation were carried out using a larger sample (and with low uncertainties) which incorporates measurements of our data and the two samples mentioned before until R$_{gal}$=25 Kpc. We find evidence that the innermost component of the NGC\,1316 GC system rotates with a PA of -10$\pm$8$\degr$ (measured from the North to the East) and a rotation velocity of 145$\pm$32 km/s. This result constitutes a new proof of the existence of an orderly movement in the inner zone of the GC system of NGC\,1316.
This behaviour is not an exception, rotation in GC systems in massive early-type galaxies has been documented in other works, for example in NGC\,4486 \citep{Cote2001}, NGC\,5128 \citep{Woodley2007,Coccato2013} and NGC\,1407 \citep{Pota2013}.

\citet{Richtler2014} measured the rotation of the diffused stellar component and obtained an PA of 72$\degr$, that is, a difference of 17$\degr$ with the major axis of the galaxy. \citet{McNeil2012} determined the rotation of the PNe associated with NGC\,1316, and obtained an PA of 64$\pm$8$\degr$. Assuming that the PNe are associated with the stellar component of NGC\,1316, we would be observing the rotation of the galaxy. In this case, the PA obtained by these authors is in agreement (within the errors) with the estimate of the major axis of the galaxy. In synthesis, this means that the PA of the GCs rotation differs by $\sim$78$\degr$ to the PA of the stellar component.

The fact that the inner GCs and the stellar components rotates almost perpendicular could indicate that NGC\,1316 is a polar-ring galaxy candidate \citep{Whitmore1990}. Several observational properties would support this idea. Among them we could mention an inner rotating disk of ionized gas \citep{Schweizer1980} and a bar-like structure observed through Spitzer images \citep{Lanz2010}, both along the minor axis of NGC\,1316. However, a more detailed study of the diffuse star component of the galaxy is necessary to arrive at a conclusion. The highly complex kinematic behaviour presented by NGC\,1316 could be a consequence of the perturbation introduced by merger events (e.g. \citealt{Bekki1998}, \citealt{Bekki&chiba2002}).


Peculiar GC kinematic profiles were analyzed by \citet{Coccato2013} for NGC\,1399, NGC\,4649 and NGC\,5128. These authors have compared the halo kinematics traced by GCs and PNe. They find differences in the rotational properties of the PNe and the different GCs subpopulations, for example, misaligned axes of rotation in NGC\,4649.

Another interesting case is NGC\,4486, the giant elliptical galaxy that dominates the Virgo cluster. \citet{Cote2001} determined that for R$_{gal}$~\textgreater~ 15\,Kpc, the subpopulation of metal-poor GCs rotate around the photometric major axis as well as the metal-rich GCs, while inside this radius, the metal-poor clusters appear to rotate around the photometric major axis.

\subsection{Integrated properties of the GCs}
\label{sec5.2}

The age and metallicity of some GCs associated with NGC\,1316 had previously been established by \citet{Goudfrooij2001a}. These authors obtained spectra for three GCs with S/N good enough to measure the H$\alpha$ (age sensitive) and CaII triplet (as an indicator of metallicity) lines. They estimated the integrated properties of the GCs by comparing these indicators with those obtained through the SSP models of Bruzual $\&$ Charlot, (1996) (obtained in a private communication). They reported that these GCs (ID$_{Goud.}$ 103, 114 and 210) had solar or subsolar abundances and ages of $\sim$3$\pm$0.5 Gyr. Only one of them was present in our spectroscopic sample, ID$_{Goud.}$=210 (ID$_{sesto}$=156). The analysis performed in this work establishes for this GC an age of 1.8$^{+0.5 }_{-0.37}$\,Gyr and [Z/H]=0.281$\pm$0.151\,dex. Our estimate of age is significantly lower than that obtained by \citet{Goudfrooij2001a}, while the metallicity that we have determined is considerably greater.

It is significant to note that the spectrum used in this work presents a considerably higher S/N value (S/N per\,\AA\,= 78), which allowed us to measure many more spectral indicators than those used by \citet{Goudfrooij2001a}. This result is particularly interesting, since in the literature this value is used to establish the age of the last merger event experienced by NGC\,1316 \citep[e.g.][]{Asabere2016, Iodice2017}.

The spectroscopic results have confirmed the photometric conclusions obtained in paper I, that is, the presence of multiple GC populations associated with NGC\,1316 was confirmed, among which stands out the presence of a population of young GCs with an average age of 2.1 Gyr and metallicities between -0.5~\textless~[Z/H]~\textless~0.5 dex.
In paper I we had measured an age of 5 Gyr and solar metallicity for this subpopulation. This result, within the errors, is in complete agreement with the new spectroscopic values. It is therefore important to emphasize that the photometric method that we had applied in paper I is very useful to separate between different subpopulations of GCs.

On the other hand, this young subpopulation presents $\alpha$-element abundances in the range -0.2~\textless~[$\alpha$/Fe]~\textless~0.3\,dex, with a slight concentration towards solar values. This is lower than the average +0.3 and +0.2\,dex [$\alpha$/Fe] ratios of GCs in the MW and M31 \citep{Puzia2005a}. Furthermore, unlike what can be observed for GCs in early-type galaxies \citep[e.g.][]{Puzia2005b} our sample does not have objects with $\alpha$-element abundances~\textgreater~0.4\,dex.

Several works have determined the age and metallicity of the stellar component in NGC\,1316 \citep[e.g.][]{Terlevich2002,Thomas2005,Silva2008,Bedregal2011}. \citet{Thomas2005} estimated the integrated properties of the stellar populations in the inner zone of the galaxy, within a galactocentric radius of 0.56\,Kpc (5.6\,arcsec). Using different Lick/IDS indices and SSP models \citep{Thomas2003}, these authors derived an age of 3.2$\pm$0.2 Gyr, [Z/H]=0.338$\pm$0.023\,dex and [$\alpha$/Fe]=0.15$\pm$0.009\,dex. \citet{Bedregal2011} estimated the integrated properties of NGC\,1316 as a function of galactocentric distance. Using Lick/IDS indices and SSP models of \citet{Vazdekis2010} these authors estimated ages, metallicities and $\alpha$-element abundances gradients along the semi-major axis of the galaxy, for distances between 1 and 8\,Kpc (10$-$80\,arcsec). The age shows a gradient with a positive slope (see Figure B1, in \citealt{Bedregal2011}) that ranges from $\sim$4 Gyr in the region closest to the galactic centre, to 10 Gyr at 8\,kpc. In this same region the metallicity presents a negative slope, covering from 0.15 to -0.7\,dex. Meanwhile, $\alpha$-element abundances remain practically constant at 0.12\,dex.
 These results are in good agreement with those obtained for the population of young GCs, which would indicate the existence of a common mechanism for the creation of both young clusters and field stars. 
This would be consistent with the scenario discussed by \citet{Whitmore2007}, in which most clusters disrupt after they form and their stars are dispersed into the field population of the galaxy.

\subsubsection{Comparison with other Stellar systems}
\label{sec5.2.1}

\begin{figure*}
	\includegraphics[width=0.9\columnwidth, angle=270]{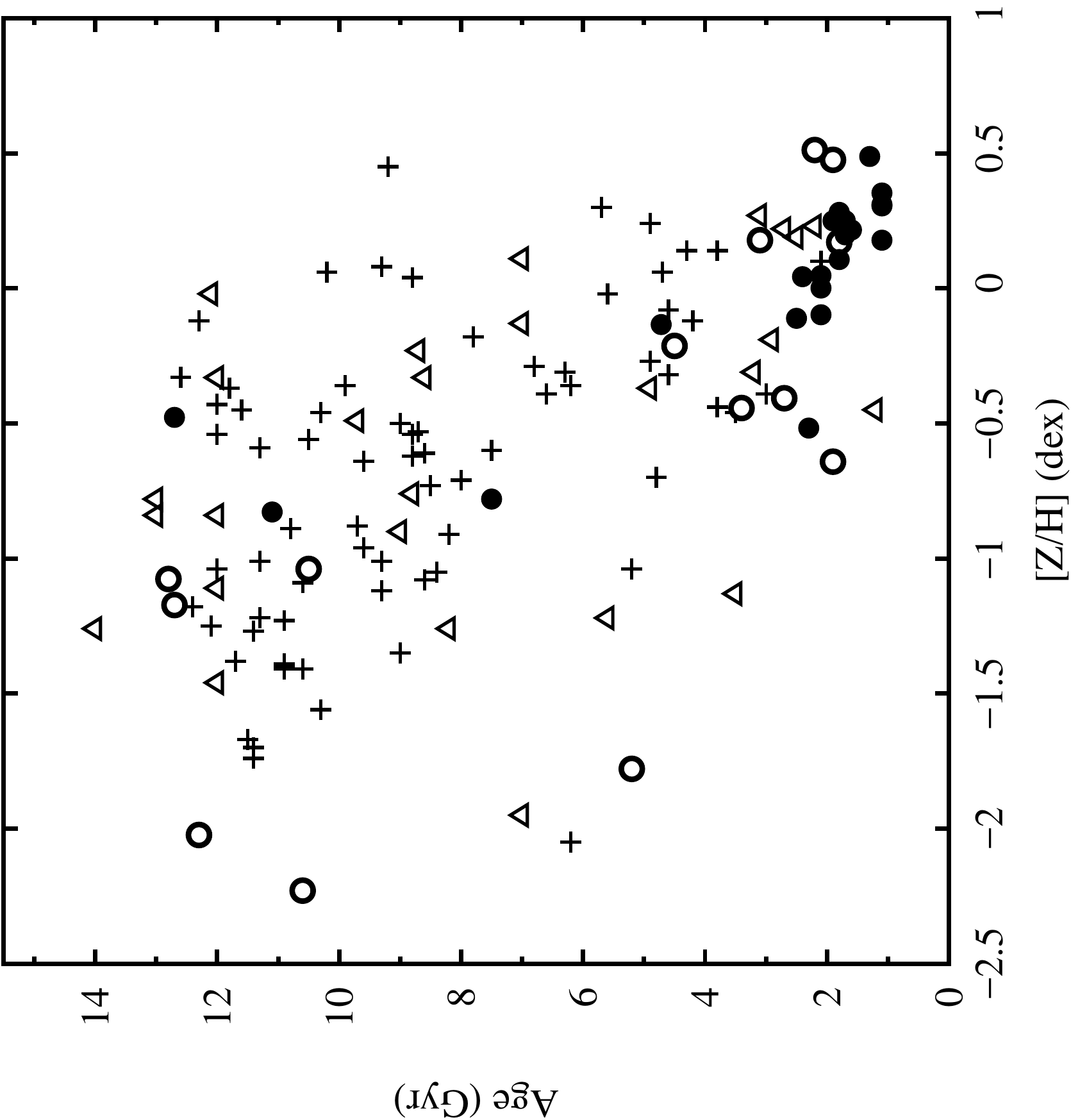}
	\includegraphics[width=0.9\columnwidth, angle=270]{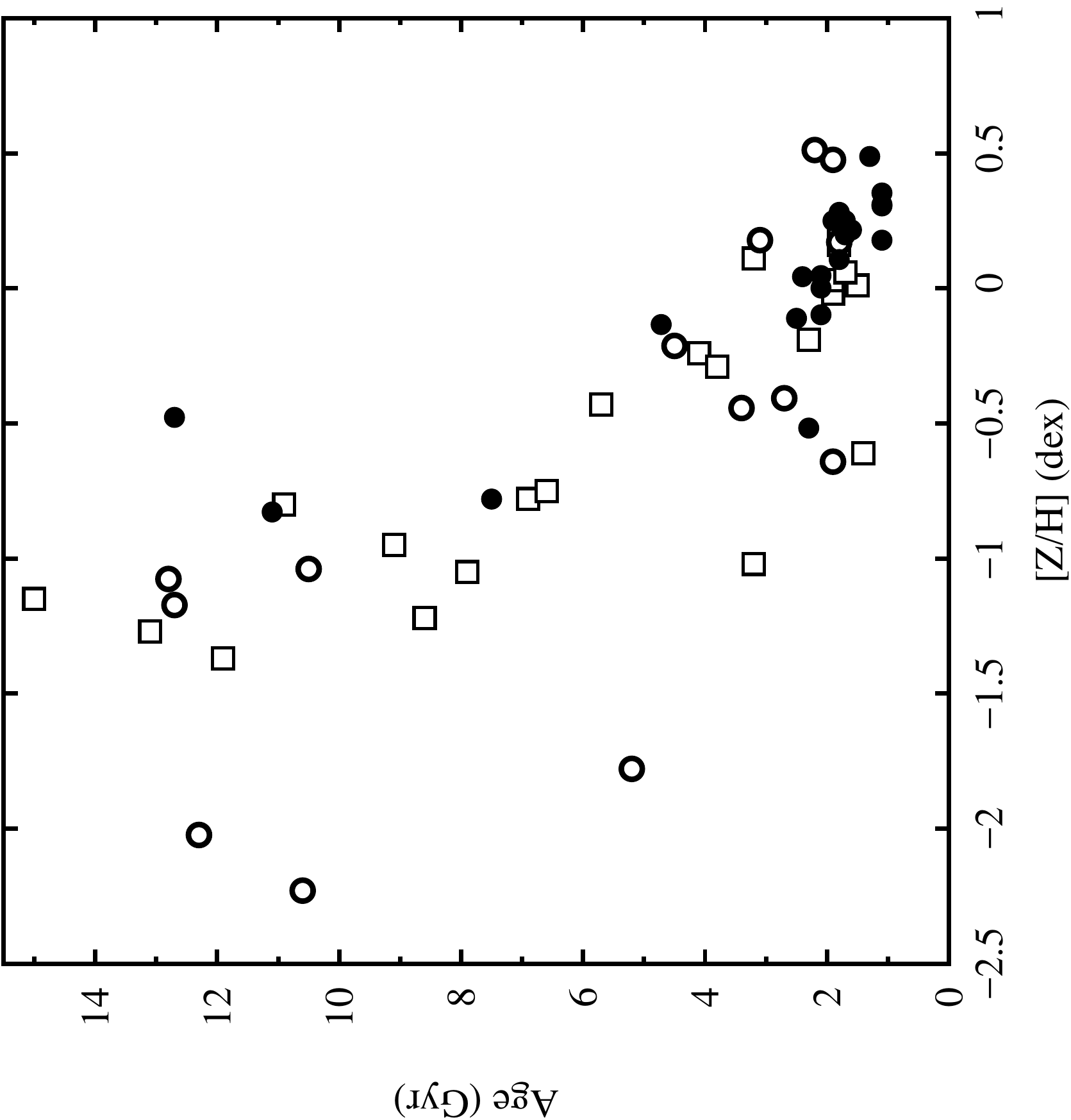}
	\includegraphics[width=0.9\columnwidth, angle=270]{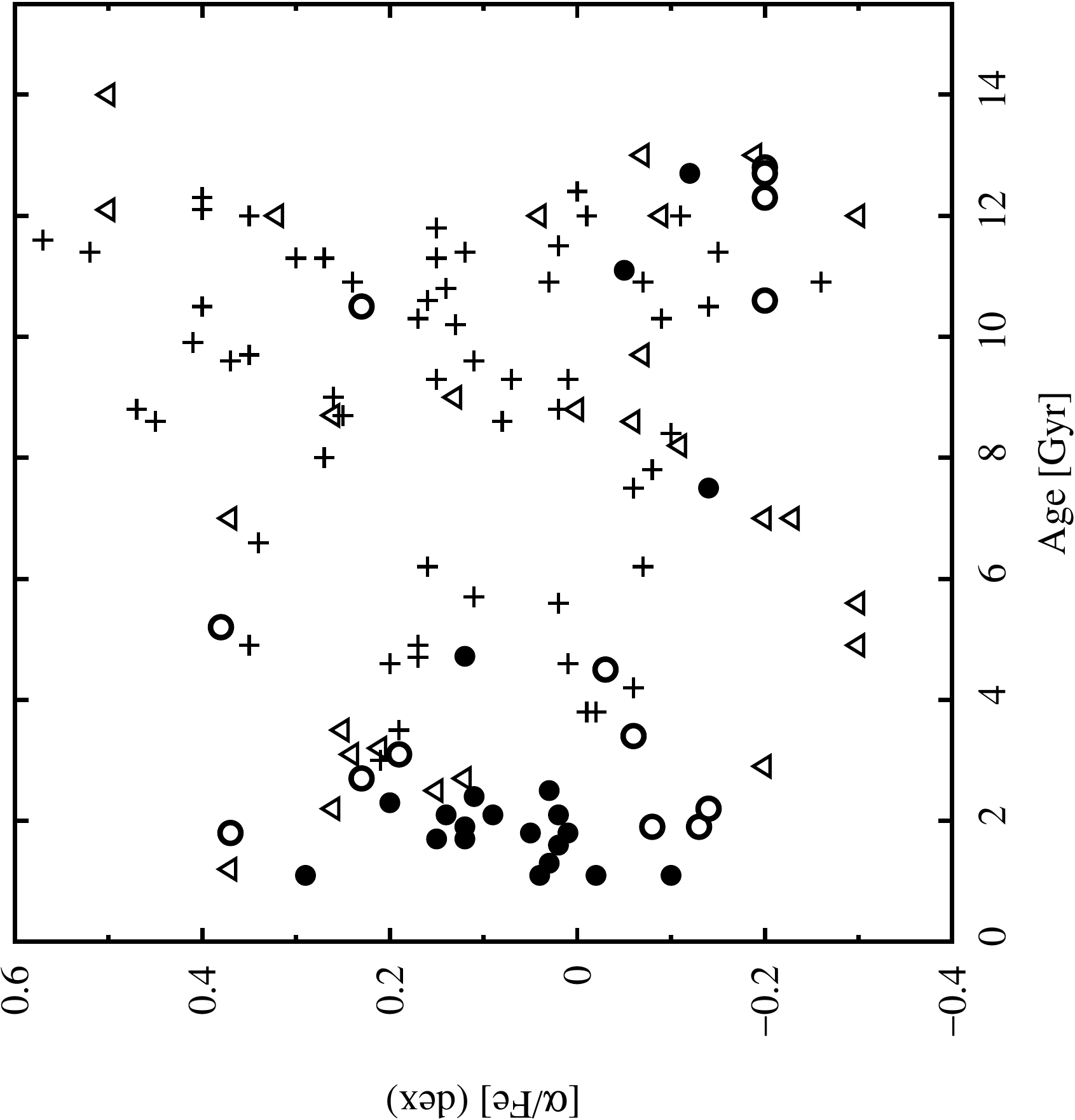}
	\includegraphics[width=0.9\columnwidth, angle=270]{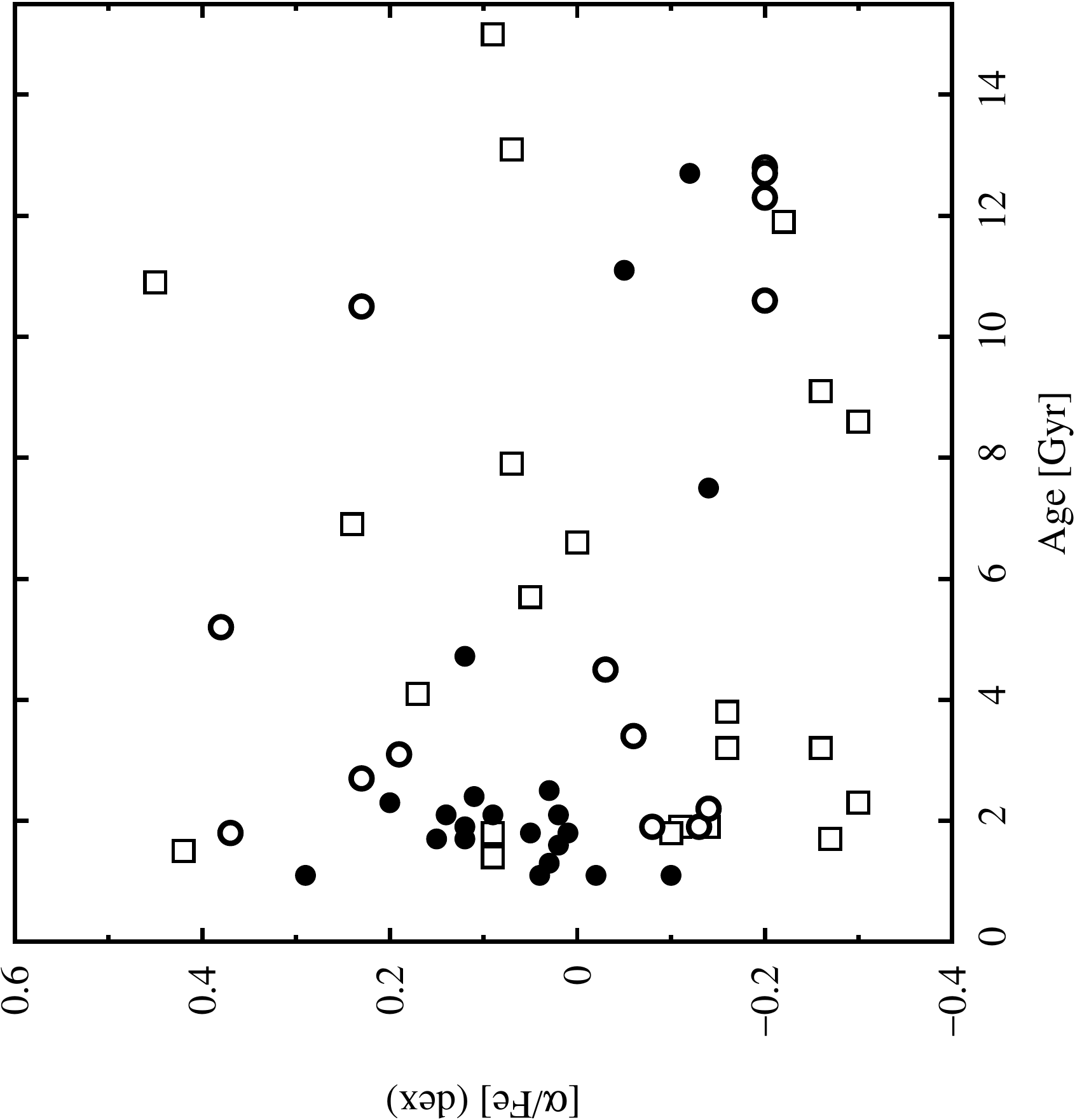}
        \includegraphics[width=0.9\columnwidth, angle=270]{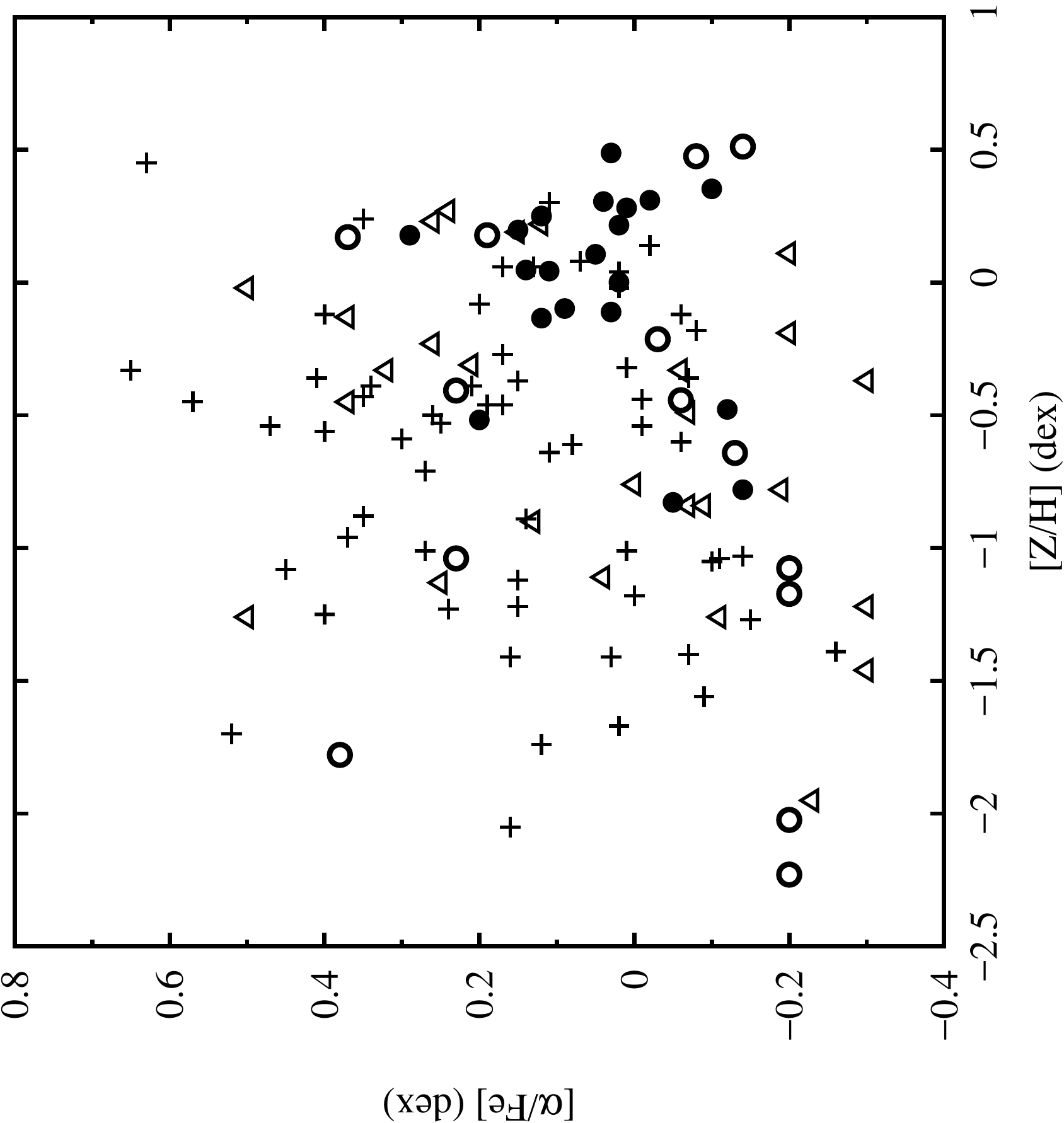}
        \includegraphics[width=0.9\columnwidth, angle=270]{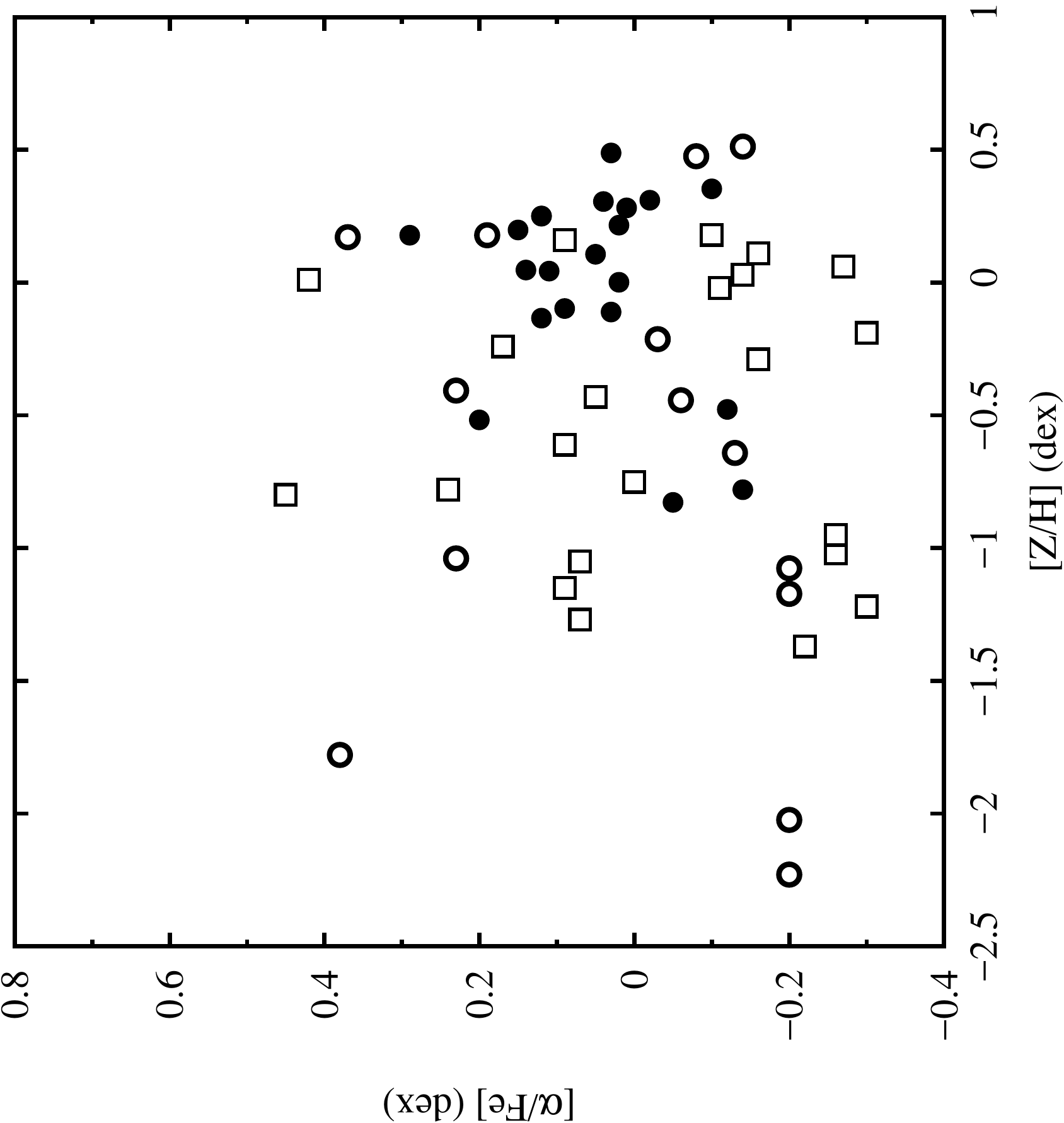}
 \caption{Left panel: Integrated properties of the GCs of the samples of NGC\,1316 (S/N~\textgreater~25 filled circles, S/N~\textless~25 empty circles), NGC\,5128 (\citealt{Woodley2010a}, plus signs) and NGC\,4636 (\citealt{Park2012}, empty triangles). Right panel:  Integrated properties of the GCs of the sample of NGC\,1316 (S/N~\textgreater~25 filled circles, S/N~\textless~25 empty circles) and nuclei of
early-type dwarf galaxies (\citealt{Paudel2011}, empty squares).}
    \label{fig22n}
\end{figure*}

Figure \ref{fig22n} shows the comparison of our results with those obtained in different works present in the literature. We used the spectroscopic results of \citet{Woodley2010a} for 72 GCs in NGC\,5128 (S/N per~\AA~\textgreater\,30) and \citet{Park2012} for 33 GCs in NGC\,4636 (S/N per~\AA~\textgreater\,15). 
 These authors have reported the presence of young and intermediate-age GCs. Furthermore, we present the comparison with 22 nuclei of early-type dwarf galaxies measured by \citet{Paudel2011} (S/N per~\AA~\textgreater\,15). All these works  obtained ages, metallicities and $\alpha$-element abundances by comparing measured Lick/IDS indices with those obtained from the SSP models of \citet{Thomas2003} and/or \citet{Thomas2004}. What is most striking in this figure is the fact that the sample of GCs of NGC\,1316 is the only one that presents a young subpopulation with very homogeneous age, metallicity and $\alpha$-element abundances. This feature is a strong argument for considering these objects as true GCs originated in the merger processes. Moreover, it is observed that there are several objects of NGC\,1316, NGC\,4636 and to a lesser extent NGC\,5128, with subsolar values of [$\alpha$/Fe] and a large spread of [Z/H] and ages. This behaviour seems to show similarities with some nuclei of early-type dwarf galaxies of \citet{Paudel2011}, as shown in the right panel of Figure \ref{fig22n}. A similar scenario is observed in Figure 6 in \citet{Janz2016}, which shows a clear separation in the [$\alpha$/Fe] values between the dwarf nuclei and GCs. 
Therefore, in this context, we cannot reject the possibility that some of the objects displaying [$\alpha$/Fe]~\textless~0 in our sample are, in fact, stripped nuclei. In order to shed light on their real nature, in a forthcoming paper we will analyze the star formation history of this particular objects.

\subsection{Possible scenario for NGC\,1316}

In the kinematic analysis we observed that the GCs present a rotation almost perpendicular to the one that was measured for the stellar component. Taking into account that the sample is dominated by a population of young GCs with relatively high metallicity, this result would be in agreement with a scenario in which the young GCs were formed from a merger with a gas-rich galaxy, conserving the angular momentum of its parent galaxy. These types of scenarios are analyzed through numerical simulations by \citet{Bekki&chiba2002}.

We may think that this galaxy has cannibalized one or more gas-rich galaxies, where the last fusion event occurred about 2 Gyr ago. In turn, there could have been numerous other minor mergers and satellite accretions. Such encounters could be the origin of the stripping events and subsequent captures of dwarf nuclei. The latter would also explain the tidal structures described by \citet{Iodice2017}.

These results, analyzed together, allow us to delineate the evolutionary history of NGC\,1316. This would indicate that this galaxy has not yet finished forming its GC system. This scenario is consistent with that published in \citet{Iodice2017}, where these authors suggest that this galaxy is still forming its halo. Thus, we could explain both the dust structures and shells. Also, due to the detection of this significant population of young GCs with characteristics similar to the stellar populations in the innermost areas of the galaxy, we know that during those processes an intense starburst occurred.

An aspect not yet very clear in this sense is the relatively low number of \textit{red} GCs in NGC\,1316. This phenomenon could be related to the progenitors that have shaped this galaxy. We tentatively suggest that the reason for this is that NGC\,1316 was originally a non-massive galaxy, dominated by \textit{blue} GCs. Then GCs of relatively high metallicity were formed, which, within several billions of years, will become a larger subpopulation of \textit{red} GCs. The evidence presented here indicates that these young GCs, and perhaps those with ages between 6 and 8 Gyr, have formed during different merger events suffered by the galaxy, since their kinematics seems to show traces of it.

With respect to the class of galaxies that have been accreted in the aforementioned fusion events, the presence of dust, enriched gas and the low rate of \textit{red} GCs would indicate that at least one of them could be a gas-rich galaxy, which possibly formed stars and GCs during the process of interaction and fusion.

\section{Main Conclusions}
\label{}

In this paper we conducted a spectroscopic study of 35 GCs associated with NGC\,1316, using the multi-object mode of the GMOS camera, mounted on Gemini South telescope. Thanks to the high quality of this data we have been able to measure, with low errors, the integrated properties of each of them, which allow us to design a possible scenario that describes the violent past of NGC\,1316. A short summary of the main conclusions is included below:

\begin{itemize}

\item Radial velocities were obtained for each of the objects of the mask, and 35 genuine GCs were confirmed. They present radial velocities close to 1760 km/s, adopted as the systemic velocity of NGC\,1316.\\

\item We obtained that object ID$_{Sesto}$=1446 is actually a field star, and not a genuine GC as it had been classified by \citet{Goudfrooij2001a}.\\

\item  We find pieces of evidence that the innermost component of the NGC\,1316 GC system rotates with a PA of -10$\pm$8$\degr$ (measured from the North to the East) and a rotation velocity of 145$\pm$32 km/s, which would indicate that the GC system rotates almost perpendicular to the stellar component.\\

\item We observed the presence of a large number of GCs brighter than $\omega$Centauri, the brightest GC-type object in the MW. Most of these objects belong to the young subpopulation with relatively high metallicity.\\

\item The sample is dominated by a population of young GCs with an average age of 2.1 Gyr, metallicities between $-$0.5~\textless~[Z/H]~\textless~0.5\,dex and $\alpha$-element abundances in the range $-$0.2~\textless~[$\alpha$/Fe]~\textless~0.3\,dex. The remarkable homogeneity presented by the integrated properties of the young objects is the strongest argument for considering these objects as bona-fide GCs.\\

\item We observed a smaller group of older GCs, with ages of 7-12 Gyr, metallicities in the range $-$1~\textless~[Z/H]~\textless~0.0\,dex, and $\alpha$-element abundances between $-$0.2 and 0.1\,dex. \\

\item There are several objects in our sample with subsolar values of [$\alpha$/Fe] and a large spread of [Z/H] and ages. This behaviour seems to show similarities with some nuclei of early-type dwarf galaxies present in the literature. We have concluded that some of our objects with [$\alpha$/Fe]~\textless~0 could actually be stripped nuclei, possibly accreted during minor merger events.\\

\item We may think that this galaxy has cannibalized one or more gas-rich galaxies, where the last fusion event occurred about 2 Gyr ago. During this process, a large number of metal-rich GCs were formed, as well as field stars. \\

\item The complex kinematic behaviour presented by NGC\,1316 and the presence of a dominant subpopulation of very young GCs could indicate that this galaxy has not yet finished forming its GC system.\\

\item This sample constitutes the broadest spectroscopic sample of young bona-fide GCs published to date.\\

\end{itemize}

\section*{Acknowledgements}

We thank the referee, B. Whitmore, for his significant contributions which greatly improved this work.

This work was funded with grants from Consejo Nacional de Investigaciones Cientificas y Tecnicas de la Republica Argentina, and Universidad Nacional de La Plata (Argentina). Based on observations obtained at the Gemini Observatory, which is operated by the Association of Universities for Research in Astronomy, Inc., under a cooperative agreement with the NSF on behalf of the Gemini partnership: the National Science Foundation (United States), the National Research Council (Canada), CONICYT (Chile), the Australian Research Council (Australia), Minist\'{e}rio da Ci\^{e}ncia, Tecnologia e Inova\c{c}\~{a}o (Brazil) and Ministerio de Ciencia, Tecnolog\'{i}a e Innovaci\'on Productiva (Argentina). 

The Gemini program ID is GS-2013B-Q-24. This research has made use of the NED, which is operated by the Jet Propulsion Laboratory, Caltech, under contract with the National Aeronautics and Space Administration.




\bibliographystyle{mnras}
\bibliography{biblio_Sesto.bib} 



\appendix

\section{Some extra material}

\begin{landscape}
\begin{table}
\footnotesize
\begin{tabular}{lcccccccccll}
\hline
\hline
\textbf{ID$_s$}&\textbf{ID$^*$}&\textbf{$\alpha_{J2000}$}&\textbf{$\delta_{J2000}$}&\textbf{g'$_0$}&\textbf{r'$_0$}&\textbf{i'$_0$}&\textbf{RV}&\textbf{[Z/H]}&\textbf{[$\alpha$/Fe]}&\textbf{Age}&\textbf{S/N}\\
$\#$&$\#$&($h:m:s$)&($\degr:\arcmin:\arcsec$)&mag&mag&mag&km/s&dex&dex&Gyr&per\,\AA\\
\hline
11&3033$^{2}$&3:22:31.73&-37:13:41.30&21.781$\pm$0.003&21.173$\pm$0.002&20.810$\pm$0.004&1426$\pm$15&0.04$\pm$0.11&0.11 $\pm$0.08&2.4$^{+1.1}_{-0.8} $&29\\
55&&3:22:38.64&-37:13:28.90&21.880$\pm$0.005&21.280$\pm$0.004&20.993$\pm$0.007&1650$\pm$22&-1.17$\pm$0.12&-0.20$\pm$0.01&12.7$^{+1.2 }_{-1.1}$&23\\
73&&3:22:36.48&-37:13:23.10&22.231$\pm$0.008&21.643$\pm$0.008&21.383$\pm$0.013&1514$\pm$26&-0.21$\pm$0.16&-0.03$\pm$0.14&4.6$^{+3.7 }_{-2.0} $&17\\
94&212$^{1}$&3:22:40.41&-37:13:18.70&21.263$\pm$0.004&20.652$\pm$0.004&20.278$\pm$0.007&1954$\pm$21&0.18$\pm$0.05&0.29 $\pm$0.08&1.1$^{+0.3 }_{-0.3} $&30\\
156&210$^{1}$&3:22:38.03&-37:13:07.00&19.790$\pm$0.003&19.139$\pm$0.004&18.778$\pm$0.006&1472$\pm$22&0.28$\pm$0.15&0.01$\pm$0.05&1.8$^{+0.5}_{-0.4} $&78\\
178&&3:22:36.70&-37:13:02.40&21.566$\pm$0.010&20.945$\pm$0.004&20.589$\pm$0.008&1915$\pm$23&0.49$\pm$0.20&0.03 $\pm$0.09&1.3$^{+0.5 }_{-0.4} $&25\\
231&111$^{1}$&3:22:37.93&-37:12:49.70&20.370$\pm$0.003&19.723$\pm$0.002&19.373$\pm$0.003&1640$\pm$22&0.11$\pm$0.2&0.05$\pm$0.05&1.8$^{+0.5}_{-0.6} $&48\\
259&112$^{1}$&3:22:39.61&-37:12:41.60&19.668$\pm$0.007&19.071$\pm$0.004&18.728$\pm$0.003&1265$\pm$20&0.20$\pm$0.20&0.15 $\pm$0.05&1.7$^{+0.6 }_{-0.4} $&85\\
295&207$^{1}$&3:22:33.92&-37:12:32.70&21.831$\pm$0.002&21.190$\pm$0.002&20.872$\pm$0.003&1619$\pm$18&-0.52$\pm$0.15&0.20 $\pm$0.16&2.4$^{+0.9 }_{-0.7} $&26\\
319&121$^{1}-$3336$^{2}$&3:22:50.86&-37:12:25.60&21.891$\pm$0.004&21.280$\pm$0.004&20.947$\pm$0.006&1546$\pm$41&-0.84$\pm$0.18&-0.11$\pm$0.11&7.5$^{+4.6}_{-2.8} $&28\\
327&205$^{1}$&3:22:31.42&-37:12:22.30&20.694$\pm$0.001&20.054$\pm$0.002&19.697$\pm$0.001&1672$\pm$13&0.31$\pm$0.22&-0.02$\pm$0.08&1.2$^{+0.4 }_{-0.3} $&69\\
351$\dagger$&&3:22:42.05&-37:12:18.0&20.384$\pm$0.038&19.336$\pm$0.017&18.944$\pm$0.015&1959$\pm$27&0.22$\pm$0.27&0.02 $\pm$0.07&1.6$^{+0.6 }_{-0.4} $&43\\
376&215$^{1}$&3:22:46.64&-37:12:10.5&21.183$\pm$0.003&20.492$\pm$0.002&20.117$\pm$0.004&1672$\pm$22&-0.13$\pm$0.15&0.12$\pm$0.06&4.7$^{+2.8}_{-1.7}$&37\\
398&&3:22:46.12&-37:12:05.0&20.131$\pm$0.011&19.495$\pm$0.011&19.123$\pm$0.013&1428$\pm$24&0.25$\pm$0.21&0.12 $\pm$0.03&1.7$^{+0.6 }_{-0.4} $&86\\
421&&3:22:46.86&-37:11:57.6&22.366$\pm$0.006&21.686$\pm$0.007&21.271$\pm$0.005&2198$\pm$28&0.17$\pm$0.26&0.37 $\pm$0.12&1.9$^{+1.6 }_{-0.9} $&16\\
457&&3:22:38.21&-37:11:49.7&22.534$\pm$0.017&22.054$\pm$0.015&21.733$\pm$0.016&1738$\pm$41&-0.64$\pm$0.27&-0.13$\pm$0.10&1.9$^{+1.1 }_{-0.7} $&13\\
485&&3:22:44.12&-37:11:42.5&22.135$\pm$0.016&21.562$\pm$0.009&21.292$\pm$0.007&1963$\pm$62&-0.44$\pm$0.26&-0.06$\pm$0.19&3.5$^{+2.5 }_{-1.5} $&15\\
500&217$^{1}-$3318$^{2}$&3:22:49.41&-37:11:38.2&21.314$\pm$0.002&20.680$\pm$0.002&20.297$\pm$0.002&1862$\pm$26&-0.11$\pm$0.11&0.03 $\pm$0.07&2.6$^{+0.8 }_{-0.6} $&44\\
515&104$^{1}$&3:22:29.94&-37:11:33.3&21.429$\pm$0.009&20.829$\pm$0.011&20.497$\pm$0.015&1990$\pm$10&0.01$\pm$0.08&0.02 $\pm$0.06&2.2$^{+0.4 }_{-0.3} $&44\\
561&115$^{1}$&3:22:43.39&-37:11:25.5&21.878$\pm$0.007&21.147$\pm$0.005&20.759$\pm$0.005&1955$\pm$27&0.35$\pm$0.22&-0.10$\pm$0.14&1.2$^{+0.4 }_{-0.3} $&29\\
636&107$^{1}$&3:22:33.25&-37:11:13.2&21.859$\pm$0.005&21.122$\pm$0.008&20.721$\pm$0.010&1412$\pm$22&-0.48$\pm$0.07&-0.12$\pm$0.08&12.8$^{+0.8}_{-0.8}$&32\\
681&3351$^{2}$&3:22:51.79&-37:11:05.1&22.191$\pm$0.003&21.616$\pm$0.004&21.330$\pm$0.004&1741$\pm$34&-2.02$\pm$0.23&-0.20$\pm$0.2&2.41$^{+4.4 }_{-3.2} $&23\\
728&110$^{1}$&3:22:36.50&-37:10:56.1&20.252$\pm$0.009&19.600$\pm$0.010&19.267$\pm$0.016&1974$\pm$11&0.05$\pm$0.07&0.14$\pm$0.05&2.12$^{+0.3}_{-0.3}$&73\\
821&&3:22:50.00&-37:10:43.2&22.372$\pm$0.004&21.892$\pm$0.005&21.581$\pm$0.007&1704$\pm$57&-1.51$\pm$0.44&0.13 $\pm$0.33&10.6$^{+4.2 }_{-3.0}$&20\\
848&106$^{1}$&3:22:31.97&-37:10:38.6&21.734$\pm$0.003&21.072$\pm$0.005&20.729$\pm$0.007&2035$\pm$28&-0.83$\pm$0.11&-0.05$\pm$0.11&11$^{+2.3}_{-1.8}$&36\\
1007&&3:22:42.59&-37:10:20.2&21.151$\pm$0.009&20.562$\pm$0.010&20.275$\pm$0.014&1773$\pm$19&0.31$\pm$0.03&0.04 $\pm$0.04&1.1$^{+0.4 }_{-0.3} $&55\\
1033&123$^{1}-$3384$^{2}$&3:22:53.85&-37:10:15.3&20.091$\pm$0.003&19.487$\pm$0.002&19.134$\pm$0.001&1991$\pm$19&-0.10$\pm$0.02&0.09 $\pm$0.02&2.2$^{+0.4 }_{-0.3} $&115\\
1169&&3:22:43.68&-37:09:59.4&23.485$\pm$0.011&22.871$\pm$0.010&22.602$\pm$0.017&1697$\pm$48&0.18$\pm$0.43&0.19 $\pm$0.31&3.1$^{+5.4 }_{-2.0} $&8\\
1220&&3:22:44.43&-37:09:51.9&22.634$\pm$0.008&22.054$\pm$0.008&21.786$\pm$0.010&2110$\pm$49&-0.70$\pm$0.64&0.00 $\pm$0.26&5.5$^{+11.5}_{-3.7} $&18\\
1265&1462$^{2}$&3:22:46.11&-37:09:44.5&22.670$\pm$0.009&21.989$\pm$0.010&21.622$\pm$0.007&1838$\pm$32&0.51$\pm$0.13&-0.14$\pm$0.07&2.3$^{+0.8 }_{-0.6} $&16\\
1322&&3:22:52.09&-37:09:36.4&23.142$\pm$0.008&22.611$\pm$0.008&22.319$\pm$0.012&1776$\pm$54&-0.41$\pm$0.57&0.23 $\pm$0.26&2.7$^{+4.6 }_{-1.7} $&11\\
1349&&3:22:47.14&-37:09:31.9&22.197$\pm$0.010&21.473$\pm$0.008&21.068$\pm$0.011&1609$\pm$29&0.25$\pm$0.07&0.12 $\pm$0.08&2.0$^{+0.5 }_{-0.4} $&26\\
1494&&3:22:37.10&-37:09:11.3&23.354$\pm$0.013&22.911$\pm$0.015&22.604$\pm$0.018&1957$\pm$48&-1.38$\pm$0.69&0.33 $\pm$0.27&5.2$^{+11.7}_{-3.6} $&9\\
1622&&3:22:35.76&-37:08:51.3&22.974$\pm$0.008&22.365$\pm$0.010&21.995$\pm$0.016&1958$\pm$20&0.48$\pm$0.18&-0.08$\pm$0.11&1.9$^{+0.5}_{-0.4} $&14\\
1789&&3:22:41.44&-37:12:55.8&22.158$\pm$0.017&21.543$\pm$0.015&21.300$\pm$0.016&1856$\pm$40&-1.08$\pm$0.17&-0.20$\pm$0.00&12.8$^{+1.3 }_{-1.2}$&15\\
\hline
&&&&&&&&&&&\\
\hline
111&&3:22:45.96&-37:13:14.70&19.970$\pm$0.002&19.425$\pm$0.001&19.163$\pm$0.001&100$\pm$22&--&--&--&114\\
1102&&3:22:39.67&-37:10:07.8&20.259$\pm$0.002&18.896$\pm$0.002&17.345$\pm$0.003&56$\pm$62&--&--&--&99\\
1446&119$^{1}$&3:22:48.60&-37:09:17.8&20.587$\pm$0.001&20.229$\pm$0.005&20.122$\pm$0.002&-38$\pm$30&--&--&--&81\\
1541&&3:22:49.67&-37:09:03.2&21.254$\pm$0.002&20.810$\pm$0.003&20.600$\pm$0.003&-12$\pm$27&--&--&--&52\\
1702&&3:22:50.59&-37:08:37.9&21.727$\pm$0.003&20.501$\pm$0.001&19.775$\pm$0.002&40$\pm$74&--&--&--&33\\
\hline
\end{tabular}
\caption{Photometric and spectroscopic data of all the objects present in the mask. ID$_S$ corresponds to the photometric sample of \citet{Sesto2016}. ID* corresponds to the sample of \citet{Goudfrooij2001a} (1) and \citet{Richtler2014} (2). The double line separates the confirmed GCs (above) from the field stars (bottom). $\dagger$: Uncorrected by reddening due to dust.}
\label{table3}
\end{table}
\end{landscape}

\bsp	
\label{lastpage}
\end{document}